\documentclass[12pt]{iopart}
\usepackage{graphicx} 
\usepackage{orcidlink}
\usepackage{amssymb}
\usepackage{acronym}
\usepackage{subcaption}
\usepackage[numbers,sort&compress,square]{natbib}
\begin{document}

\acrodef{LIGO}{Laser Interferometer Gravitational-wave Observatory}
\acrodef{DetChar}{Detector Characterization}
\acrodef{GW}{gravitational wave}
\acrodefplural{GW}[GWs]{gravitational waves}
\acrodef{LVK}{LIGO-Virgo-KAGRA}
\acrodef{NS}{neutron star}
\acrodefplural{NS}[NSs]{neutron stars}
\acrodef{CW}{continuous wave}
\acrodef{SGWB}{stochastic gravitational wave background}
\acrodef{SFT}{short Fourier transform}
\acrodefplural{SFT}[SFTs]{short Fourier transforms}
\acrodef{PSD}{power spectral density}
\acrodef{CSD}{cross spectral density}
\acrodef{GPS}{Global Positioning Satellite}

\title[Narrow spectral artifact investigations in LIGO O4 data]{Narrow spectral artifact investigation and mitigation in LIGO data from the fourth LIGO-Virgo-KAGRA observing run}

\author{
E~Goetz\,\orcidlink{0000-0003-2666-721X}$^1$, 
A~Neunzert\,\orcidlink{0000-0003-0323-0111}$^2$, 
A~M~Knee\,\orcidlink{0000-0003-0703-947X}$^{1,3}$, 
A~Calafat\,\orcidlink{0009-0008-7515-6305}$^4$, 
X~Fan\,\orcidlink{0009-0001-9329-4065}$^5$, 
J~R~M\'erou\,\orcidlink{0000-0002-5776-6643}$^4$, 
K~A~Pham\,\orcidlink{0000-0002-7650-1034}$^6$, 
T~Starkman\,\orcidlink{0000-0003-3583-3742}$^{1,7,8}$, 
N~Aggarwal$^{9}$, 
Z~Bhalla$^{10}$,
P~Baxi$^3$, 
J~Bayley\,\orcidlink{0000-0003-2306-4106}$^{11}$, 
Y~Bu$^{12}$, 
J~B~Carlin\,\orcidlink{0000-0001-5694-0809}$^{12}$, 
P~Charlton\,\orcidlink{0000-0002-4263-2706}$^{13}$,
X~Chen$^{14}$, 
G~Cheng$^5$, 
T~Cheunchitra\,\orcidlink{0009-0001-2292-1914}$^{12}$, 
N~Christensen\,\orcidlink{0000-0002-6870-4202}$^{15,10}$, 
A~Claveus$^{16}$, 
C~M~Compton$^2$, 
M~W~Coughlin\,\orcidlink{0000-0002-8262-2924}$^6$,
F~De~Lillo\,\orcidlink{0000-0003-4977-0789}$^{17}$,
L~Dunn\,\orcidlink{0000-0002-1769-6097}$^{12}$, 
S~E~Dwyer$^{2}$,
A~Effler\,\orcidlink{0000-0001-8242-3944}$^{18}$, 
T~A~Ferreira$^{19}$
% ...
R~Weiss\footnote{Deceased, August 2025}$^{37}$, 
C~Whittle\,\orcidlink{0000-0002-8833-7438}$^{34}$, 
B~Willke\,\orcidlink{0000-0003-0524-2925}$^{41,42}$, 
C~C~Wipf$^{34}$, 
J~L~Wright$^{51}$, 
V~A~Xu\,\orcidlink{0000-0002-3020-3293}$^{30}$, 
H~Yamamoto\,\orcidlink{0000-0001-6919-9570}$^{34}$, 
L~Zhang$^{34}$, 
M~E~Zucker$^{37,34}$, 
}

\address{$^1$ University of British Columbia, Vancouver, BC V6T 1Z4, Canada}
\address{$^2$ LIGO Hanford Observatory, Richland, WA 99352, USA}
\address{$^3$ University of Michigan, Ann Arbor, MI 48109, USA}
\address{$^4$ IAC3--IEEC, Universitat de les Illes Balears, E-07122 Palma de Mallorca, Spain}
\address{$^5$ University of Chinese Academy of Sciences / International Centre for Theoretical Physics Asia-Pacific, Beijing 100190, China}
\address{$^6$ University of Minnesota, Minneapolis, MN 55455, USA}
\address{$^7$ University of Washington Bothell, Bothell, WA 98011, USA}
\address{$^8$ University of Pittsburgh, Pittsburgh, PA 15260, USA}
\address{$^{9}$ University of California, Davis, Davis, CA 95616, USA}
\address{$^{10}$ Carleton College, Northfield, MN 55057, USA}
\address{$^{11}$ IGR, University of Glasgow, Glasgow G12 8QQ, United Kingdom}
\address{$^{12}$ OzGrav, University of Melbourne, Parkville, Victoria 3010, Australia}
\address{$^{13}$ OzGrav, Charles Sturt University, Wagga Wagga, New South Wales 2678, Australia}
\address{$^{14}$ OzGrav, University of Western Australia, Crawley, Western Australia 6009, Australia}
\address{$^{15}$ Universit\'e C\^ote d'Azur, Observatoire de la C\^ote d'Azur, CNRS, Artemis, F-06304 Nice, France}
\address{$^{16}$ St.~Thomas University, Miami Gardens, FL 33054, USA}
\address{$^{17}$ Universiteit Antwerpen, 2000 Antwerpen, Belgium}
\address{$^{18}$ LIGO Livingston Observatory, Livingston, LA 70754, USA}
\address{$^{19}$ Instituto Nacional de Pesquisas Espaciais, 12227-010 S\~{a}o Jos\'{e} dos Campos, S\~{a}o Paulo, Brazil}
% ...
\address{$^{30}$ University of California, Berkeley, CA 94720, USA}
% ...
\address{$^{34}$LIGO Laboratory, California Institute of Technology, Pasadena, CA 91125, USA} 
% ...
\address{$^{37}$LIGO Laboratory, Massachusetts Institute of Technology, Cambridge, MA 02139, USA} 
% ...
\address{$^{41}$Max Planck Institute for Gravitational Physics (Albert Einstein Institute), D-30167 Hannover, Germany} 
\address{$^{42}$Leibniz Universit\"{a}t Hannover, D-30167 Hannover, Germany} 
% ...
\address{$^{51}$OzGrav, Australian National University, Canberra, Australian Capital Territory 0200, Australia} 
% ...

\ead{evan.goetz@ligo.org}

\begin{abstract}
    We present efforts to identify, characterize, and mitigate narrow spectral artifacts in LIGO detector data during the fourth LIGO-Virgo-KAGRA observing run.
    Narrow spectral artifacts in gravitational-wave detectors are non-astrophysical noise sources that can degrade searches for narrowband persistent gravitational waves.
    Identifying and, where possible, mitigating these noise sources is one of the core efforts of the LIGO Detector Characterization group.
    Key software tools have been updated and new tools deployed for the fourth LIGO-Virgo-KAGRA observing run to facilitate investigations and data monitoring.
    We discuss these tool upgrades, and present several identified narrowband artifacts that have been successfully investigated and mitigated in LIGO data.
    Regardless of whether artifacts are mitigated or not, narrowband persistent gravitational-wave searches require information on which frequency bands contain non-astrophysical artifacts.
    Minimizing the number of bands containing non-astrophysical artifacts is essential to maximize the potential for discovery of a new class of gravitational-wave signals.
\end{abstract}

\submitto{\CQG}

\maketitle

\section{\label{sec:intro}Introduction}


The \ac{LIGO}~\citep{Aasi2015} is part of a worldwide set of kilometer-scale \ac{GW} detectors forming the \ac{LVK} network.
\Ac{LIGO}, Virgo~\citep{Acernese_2015}, and KAGRA~\citep{10.1093/ptep/ptaa125} work jointly to observe astrophysical \ac{GW} signals and maximize the potential for scientific discovery.
These ground-based detectors are subject to a variety of external and detector noise sources which can diminish detector sensitivity to astrophysical \ac{GW} sources or obscure putative signals.
Scientists, engineers, and technicians work together to minimize the coupling of external and detector noise sources to the primary data output of the detectors: the time-series of measured strain, $h(t)$.
This effort presents a substantial challenge, often requiring continual efforts to find and identify new noise sources and reduce or mitigate their impact on $h(t)$.
In addition, as detectors evolve and become more sensitive (often with the installation of new hardware), new noise sources are discovered requiring further noise identification and mitigation efforts.

The LIGO \ac{DetChar} group plays a crucial role in these noise investigation and mitigation efforts~\citep{Soni_2025}.
Noise artifacts that interfere with detector sensitivity, searches for \ac{GW} signals, and estimates of signal parameters are generally classified as follows:
short duration artifacts, commonly referred to as ``glitches'', are typically short duration ($\lesssim$few seconds) and often broad in frequency;
long duration, broadband artifacts generally degrade the broad sensitivity of detectors;
and long duration, narrowband artifacts (possibly slowly evolving in frequency or amplitude), commonly referred to as ``lines'', are persistent and narrow in frequency.

Lines are particularly problematic for narrowband persistent \ac{GW} searches, while transient artifacts generally have little impact on these searches.
Prototypical narrowband persistent \ac{GW} sources are rapidly-rotating, non-axisymmetric \acp{NS}~\citep{RilesLivingReviewCW}.
The signal amplitude is expected to be weak ($h_0 < 10^{-24}$), persistent for long time scales, and nearly monochromatic.
Since detectors are located on Earth, the signal in $h(t)$ will be Doppler modulated and amplitude modulated as the Earth rotates on its axis and orbits the Sun.
Nearly monochromatic, persistent \acp{GW} sources are referred to as \ac{CW} sources.

Although their impact is most direct for \ac{CW} searches, lines can also degrade cross-correlation based analyses such as \ac{SGWB} searches, which aim to detect a weak, broadband superposition of unresolved \ac{GW} signals~\citep{Meyers:2018nyo}. 
These analyses operate at comparatively coarse frequency resolution where each frequency bin represents an average over many shorter segments. 
As a result, narrow spectral features, particularly if they are persistent or have sufficient amplitude, can accumulate power within individual frequency bins leading to non-negligible contributions to the cross-correlation measurements.
Directional (anisotropic) \ac{SGWB} searches, which map \ac{GW} power on the sky without assuming a specific waveform model, can further exhibit sensitivity to persistent features in the data in a more model-agnostic manner. 
Narrowband cross-correlation analyses utilizing the same framework to search in a model-agnostic fashion for narrowband \ac{GW} signals are additionally impacted by lines in the same manner.
Identifying and characterizing lines is therefore important for both narrowband and broadband \ac{SGWB} searches.

\begin{figure}
    \centering
    \includegraphics[width=1\textwidth]{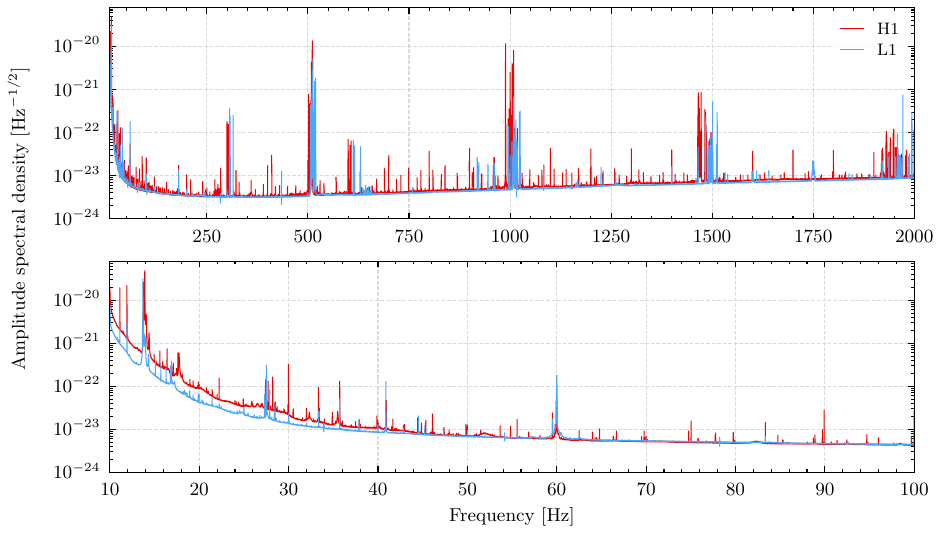}
    \caption{High-resolution, O4 run-averaged spectra of strain data collected by the \ac{LIGO} Hanford and \ac{LIGO} Livingston detectors. Upper panel: spectra from 10-2000 Hz; lower panel: zoom-in view of the low-frequency portion of the spectrum from $10$ to $100$ Hz. These spectra were calculated from $7200$-second self-gated \acsp{SFT} (frequency resolution ${\approx}0.139$~mHz), omitting time segments with poor data quality (CAT1 vetoes).}
    \label{fig:O4spectra}
\end{figure}

Narrowband persistent spectral artifacts have been studied in ground-based \ac{GW} detector data for over 20~years~\citep{PhysRevD.72.102004}.
Since 2015, Advanced \ac{LIGO} detector data from the first two observing runs (O1 and O2) included many noise sources, several of which have been mitigated, while others were not mitigated~\citep{PhysRevD.97.082002}.
Here we make use of the most recent \ac{LIGO} data from the fourth \ac{LVK} observing run (O4), which ran 24 May 2023 15:00 UTC to 18 Nov 2025 16:00 UTC, to evaluate and improve detector data through investigation and, ideally, mitigation efforts. 
O4 is longer and the strain data are more sensitive than all previous Advanced \ac{LIGO} observing runs (O1-O3) combined.
High-resolution spectra of strain data from the \ac{LIGO} Hanford and \ac{LIGO} Livingston detectors, averaged over the entire O4 run, are shown in \fref{fig:O4spectra}.

\begin{figure}
    \centering
    \includegraphics[width=1\textwidth]{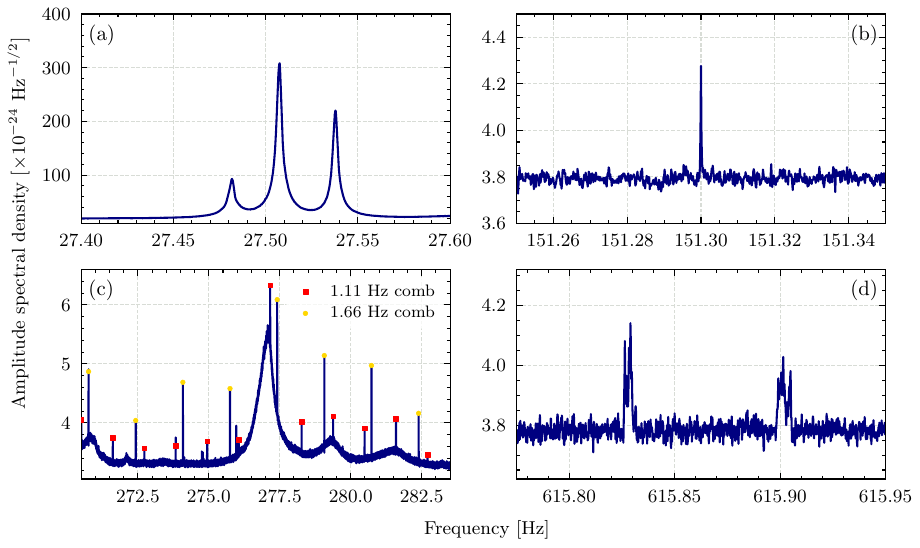}
    \caption{Examples of spectral artifacts commonly found in \acs{LIGO} $h(t)$ data: (a) broad lines resulting from optic suspension resonances; (b) a narrow line without an identified source; (c) two sets of comb artifacts with different frequency spacings; (d) lines that evolve in frequency over time, producing multi-peaked structures in run-averaged spectra. The two artifacts shown in panel (d) are part of a larger comb family with $9.5$~Hz spacing.}
    \label{fig:line_types}
\end{figure}

The spectra shown in \fref{fig:O4spectra} reveal a multitude of persistent line artifacts present in $h(t)$ data.
Lines can have several characteristics: 
1) \ac{CW} searches are particularly concerned with narrow lines, usually spanning less than $\sim$few~mHz.
Lines larger than $\sim$few~mHz are referred to as ``broad lines'', while lines $\lesssim$1~mHz are referred to as ``narrow lines''.
Although broad lines may not mimic a \ac{CW} signal in the same way as a narrow line, they can obscure potential signals and degrade sensitivity the same way as broadband noise.
Broad lines can also degrade narrowbad \ac{SGWB} searches.
2) Lines may be fixed in frequency for long duration or may slowly drift in frequency in a narrow band, typically less than $\sim$0.1~Hz.
3) Lines may be fixed in amplitude or have slow amplitude variation.
4) Lines that are separated by a common difference in frequency are known as ``combs'', while lines that are not part of a comb are individual lines.
5) Lines that belong to a comb may be observed as double, triplet, or other multi-line-like features.
6) Several individual lines that impact a small frequency band ($\sim$few~Hz) and are part of a group of lines that have common time variation may be caused by a common source.
7) Line frequency or amplitude can be impacted by detector configuration changes if the change impacts the source or the coupling mechanism.
When these change points are observed, they can be key indicators for line identification and mitigation efforts.
Examples of different line artifacts seen in the \ac{LIGO} detectors during O4 are shown in \fref{fig:line_types}.

Narrow lines and combs whose origin is known to be non-astrophysical are cataloged to enable \ac{CW} searches to veto unusable data or search candidates due to the presence of lines.
These catalogs are a key data quality product used by persistent \ac{GW} searches~\citep{Soni_2025}.
Investigations of lines and combs will be discussed in detail in \sref{sec:case-studies} and the cataloging process in \sref{sec:line-lists}.
During O4, some sources of artifacts were known to be on or off at specific times.
This information, where known, is also recorded in the catalog.

Besides the $h(t)$ channel, \ac{GW} detectors collect and record data from many other interferometric sensor and physical-environment sensors, collectively known as auxiliary channels.
These channels are essential for low-noise, high-sensitivity operations of the interferometric detector and to monitor the local environment for physical effects that may impact the detector.
Auxiliary channels are also crucial for identifying noise sources and coupling mechanisms, and they aid in proving that signal power seen at a particular frequency is due to a non-astrophysical effect.
We cautiously approach labeling spectral features as non-astrophysical artifacts in order to minimize the false-dismissal of a true narrowband persistent astrophysical \ac{GW} signal.

In addition to high-resolution line catalogs used in \ac{CW} searches, a complementary data quality product for \ac{SGWB} analyses are notch lists, which catalog frequency ``notches'' excluded from cross-correlation measurements.
The list includes frequency regions impacted by instrumental artifacts observed in the \ac{CSD}, statistically significant coherence outliers, and regions exhibiting large temporal variability throughout the observing run. 
Notch lists are discussed further in \sref{sec:Notch_lists}.

This article is organized as follows. 
In \sref{sec:methods} we describe software tools and techniques used for data quality monitoring and understanding artifacts. 
\Sref{sec:case-studies} demonstrates different case studies of investigations and mitigation efforts during O4.
In \sref{sec:dq-products}, we summarize how data quality products are generated for use by narrowband persistent \ac{GW} searches.
Finally, in \sref{sec:future} we discuss the outlook for future efforts in line investigation and mitigation in \ac{GW} detectors.

\section{\label{sec:methods}Methods}
Several software programs exist for monitoring $h(t)$ and auxiliary channel data for lines~\citep{PhysRevD.97.082002}.
Among these, Fscan~\citep{goetz_2025_18776397} is employed by the \ac{LIGO} \ac{DetChar} group for this purpose because investigation of narrow spectral artifacts generally requires high-resolution frequency-domain data (e.g. $\leq$1~mHz resolution).
Complementary tools, such as STAMP-PEM~\citep{Meyers:2018nyo} and StochMon~\citep{Stochmon}, are used for lower frequency-resolution (0.1~Hz) and are tuned for cross-correlation analyses like \ac{SGWB} searches.

\subsection{Fscan overview}

Fscan produces normalized and averaged spectra, spectrograms, and, more recently, coherence between auxiliary channels and $h(t)$.
These products are produced on a daily, weekly, and monthly basis in order to better understand line behavior that may evolve on a daily basis, or to find lines that are particularly low in amplitude and need long integration times to increase their signal-to-noise ratio.

Like other \ac{DetChar} software tools that analyze data from multiple channels at the same time~\citep{Soni_2025}, Fscan utilizes the HTCondor high-throughput computing infrastructure~\citep{htcondor_team_2024_14238998} and runs on the \ac{LIGO} computing clusters located at \ac{LIGO} Hanford and \ac{LIGO} Livingston.
Unlike other tools, however, because the cadence of the results are only as fast as daily results, we do not require these results at low-latency.
Instead, we only require that the results are available within a reasonable period of time for each step of the cadence.

Fscan has been used in some form since the initial \ac{LIGO} era in the early 2000s.
Rudimentary line counting methods in the $h(t)$ channel have been developed and employed during the initial era.
In O3, the ability to produce coherence with auxiliary channels was introduced as a new feature.
By the end of O3, however, it was decided to retire the software in its form because it was difficult to develop new features, maintain robust operations, and it did not integrate into other \ac{DetChar} software tools easily.

Prior to the start of O4, Fscan was completely rewritten as a Python-based software package to address these shortcomings~\citep{goetz_2025_18776397}.
Besides maintaining the features of Fscan employed in O3, several new features were added as part of this overhaul:
1) A line persistency figure-of-merit based on the $Z$-statistic~\citep{enwiki:1351281889} was added to better track the time-dependent nature of lines.
2) Output webpages were rebuilt to enable integration into the daily \ac{LIGO} Summary Pages.
3) The high resolution spectral data was saved in Bokeh~\citep{Bokeh} plots to enable interactive plots and better data visualization.
4) An improved line counting algorithm was employed based on a topographic prominence algorithm.
5) User interface improvements to show key figures-of-merit for at-a-glance assessment of data quality.

\subsection{\label{sec:fscan-monitor}Data monitoring with Fscan}

Data for a given detector data channel ($h(t)$ or an auxiliary channel) over a certain observing period, with duration denoted by $T_{\rm avg}$, is divided into short intervals, with duration denoted by $T_{\rm SFT}$ (typically $T_{\rm SFT} = 1800$~s).
\Ac{SFT} data is computed using the LALSuite~\citep{lalsuite} program lalpulsar\_MakeSFTs for each $T_{\rm SFT}$ where the time domain data is Hann windowed and overlapping by 50\%.
Except for instances where only physical environment monitor channels are used for a particular analysis, only times where the detector is in nominal low-noise configuration is used.

Averaging intervals are typically 1~day, 1~week, or 1~month, where daily averaging occurs over each day starting at 00:00:00 UTC, weekly averaging occurs over each week starting at 00:00:00 Wednesday UTC, and monthly averaging occurs over each month starting on the first of the month at 00:00:00 UTC.
Daily averaging allows for regular checks for changes that may occur on short time scales while weekly and monthly averaging allow for study of weaker line artifacts that may not be visible in daily spectra.
An important change for O4 was aligning the weekly average to start on Wednesdays instead of Sundays in order to more easily monitor for changes that may have occurred during weekly maintenance that happens on Tuesdays.

\Ac{SFT} data is processed using LALSuite programs to create spectral averages, spectrograms, coherence, and persistency results.
Plots and webpages of these results are created using the Fscan package to visualize and make the results accessible to \ac{LVK} scientists.

\subsubsection{Spectral averages} \label{sec:spectral_averages}
Two different types of spectral averaging on a given channel is carried out using the \ac{SFT} data.
First, normalized, averaged spectra are computed via
\begin{equation}\label{eq:normspect}
    \bar{P}(f_n) = \frac{2}{T_{\rm SFT}} \frac{1}{M} \sum_{m=0}^{M-1}\frac{|\tilde{x}_m(f_n)|^2}{\mathrm{E}[|\tilde{x}_m(f_n)|^2]}
\end{equation}
where $\tilde{x}_m(f_n)$ is the complex-valued Fourier coefficient at frequency bin $n$ of \ac{SFT} index $m$, $M$ is the total number of \acp{SFT} in an averaging interval, and $\mathrm{E}[\,]$ is the expected power, which is computed from a running median over a specified number of frequency bins.
This aids data visualization by smoothing out broader noise variations which enhances visibility of narrow spectral features.
The running median is typically computed over a window of 101~bins, $\approx$56.1~mHz.

The second method is an alternative weighting scheme, more closely aligned with that of broadband, all-sky \ac{CW} searches~\citep{o4a_allsky_cw}.
This method effectively weights frequency bins entering the weighted sum based on the variance of the data in a given \ac{SFT}.
This is defined as
\begin{equation} \label{eq:noiseweightedspect}
    \langle P(f_n)\rangle = \frac{2}{T_{\rm SFT}} \frac{\sum_{m=0}^{M-1}w_m(f_n)|\tilde{x}_m(f_n)|^2}{\sum_{m=0}^{M-1}w_m(f_n)}
\end{equation}
where the weights are defined as
\begin{equation}
    w_m(f_n) = \frac{K+1}{ \sum_{i=-K/2}^{K/2} |\tilde{x}_m(f_i)|^2 }\,.
\end{equation}
Here $K$ defines the number of bins used to establish the background estimator for frequency bin $n$ and is typically 21 bins, $\approx$11.7~mHz.

\begin{figure}
    \centering
    \includegraphics[width=1\textwidth]{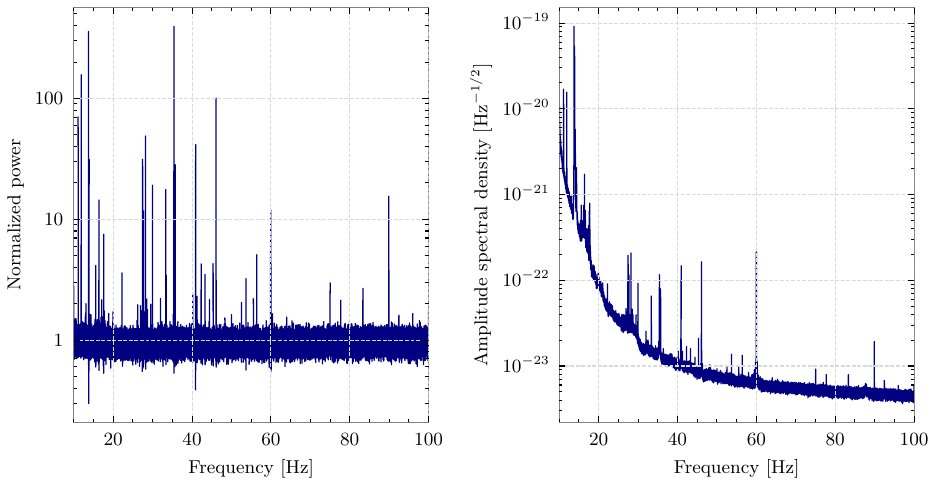}
    \caption{Example spectra produced by averaging all of the low-noise, high-sensitivity $h(t)$ data from \ac{LIGO} Hanford detector on 1 Jan 2025 00:00 UTC to 2 Jan 2025 00:00 UTC. Left panel: normalized spectral power data according to \eref{eq:normspect}. Right panel: noise-weighted amplitude spectral data according to \eref{eq:noiseweightedspect}. In both cases, $T_{\rm SFT}=1800$~s.}
    \label{fig:fscan-spectra}
\end{figure}

Examples of both spectral averages are shown in \fref{fig:fscan-spectra} over one day of data.
This inverse-variance weighting is used to produce the run-averaged spectra in \fref{fig:O4spectra} and \fref{fig:line_types}.

\subsubsection{Spectrograms}
\begin{figure}
    \centering
    \includegraphics[width=1\textwidth]{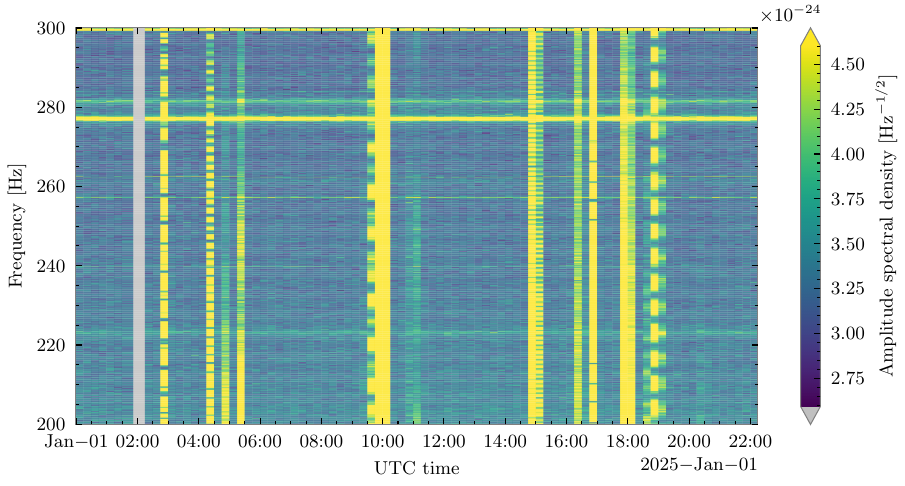}
    \caption{Example spectrogram produced from the low-noise, high-sensitivity $h(t)$ data from \ac{LIGO} Hanford detector on 1 Jan 2025 00:00 UTC to 2 Jan 2025 00:00 UTC. Note that the grey band indicates a gap in the low-noise, high-sensitivity data and bright vertical bands are short duration glitches that contribute to broadband power in the \ac{SFT}. Bright horizontal bands are narrowband persistent spectral artifacts.}
    \label{fig:fscan-spectrogram}
\end{figure}

Spectra are useful for evaluating the average frequency content over a given interval, but spectrograms provide information about spectral changes over time.
For each averaging interval, spectrograms are created from the \acp{SFT} available for that time interval.
These spectrograms are not spectrograms in the traditional sense, i.e., following Welch's method.
Instead, the power in each \ac{SFT} is calculated and these are simply plotted as power versus time.
When there are gaps in the low-noise, high-sensitivity $h(t)$ data and no \acp{SFT} are produced, a gap in the spectrogram is indicated as a gray band.
The \ac{SFT} powers in each frequency bin are also typically coarse-grained by averaging neighbouring frequency bins of width typically 0.1~Hz.
An example spectrogram is shown in \fref{fig:fscan-spectrogram}.

\subsubsection{\label{sec:coherence}Coherence}
Tracing the source of non-astrophysical narrowband spectral artifacts in $h(t)$ can be challenging because auxiliary channels may be noisy with many narrowband spectral artifacts that spuriously correlate with a narrowband artifact in $h(t)$.
Computing the coherence between $h(t)$ and an auxiliary channel is a more robust and effective metric than correlation of power alone.
Over a given averaging interval, the coherence between two channels is computed as
\begin{equation}
C(f_n) = \frac{\sum_{m=0}^{M-1}|\tilde{x}_m(f_n)\tilde{y}_m^\ast(f_n)|^2}{\sum_{m=0}^{M-1}|\tilde{x}_m(f_n)|^2\sum_{m=0}^{M-1}|\tilde{y}_m(f_n)|^2}
\end{equation}
where $\tilde{x}$ and $\tilde{y}$ denote the complex Fourier coefficients of two different channels and $\ast$ denotes the complex-conjugate.
Coherence with one or more auxiliary channels may help localize a source of an artifact, but the coupling mechanism to $h(t)$ may not necessarily be understood from the coherence alone.

\begin{figure}
    \centering
    \includegraphics[width=1\textwidth]{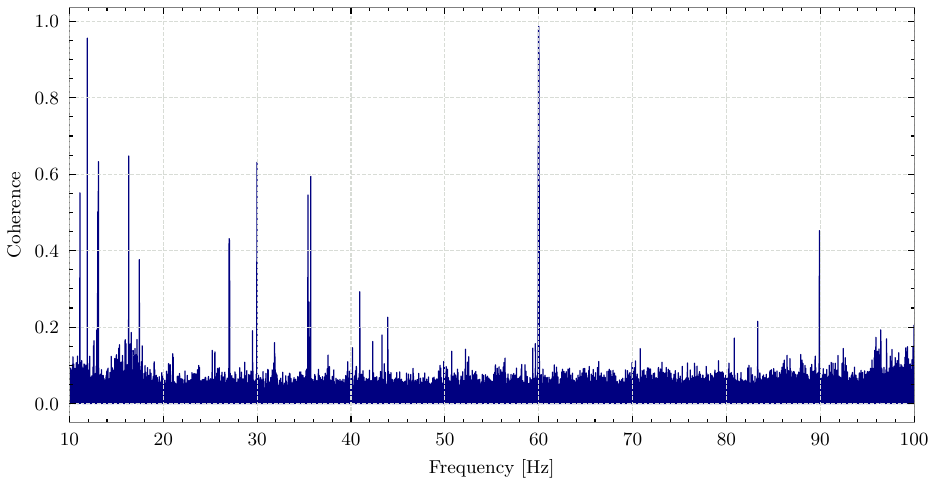}
    \caption{Example coherence spectrum calculated between the low-noise, high-sensitivity $h(t)$ data and the channel that monitors the mode cleaner length (\texttt{H1:LSC-MCL\_IN1\_DQ}) from \ac{LIGO} Hanford detector on 1 Jan 2025 00:00 UTC to 2 Jan 2025 00:00 UTC.}
    \label{fig:fscan-coherence}
\end{figure}

Fscan computes high-resolution coherence spectra of $\sim$75 auxiliary channels and $h(t)$ per detector, although the only limitation is due to use of computing and data storage resources.
These channels are chosen for their sensitivity to detector artifacts and the environmental conditions around the detectors.
As an example, the coherence between the low-noise, high-sensitivity $h(t)$ data and the mode cleaner length monitor channel (\texttt{H1:LSC-MCL\_IN1\_DQ}) for 1 Jan 2025 00:00 UTC to 2 Jan 2025 00:00 UTC is shown in \fref{fig:fscan-coherence}.

\subsubsection{\label{sec:persistFOM}Persistency figure-of-merit}
\begin{figure}
    \centering
    \includegraphics[width=1\textwidth]{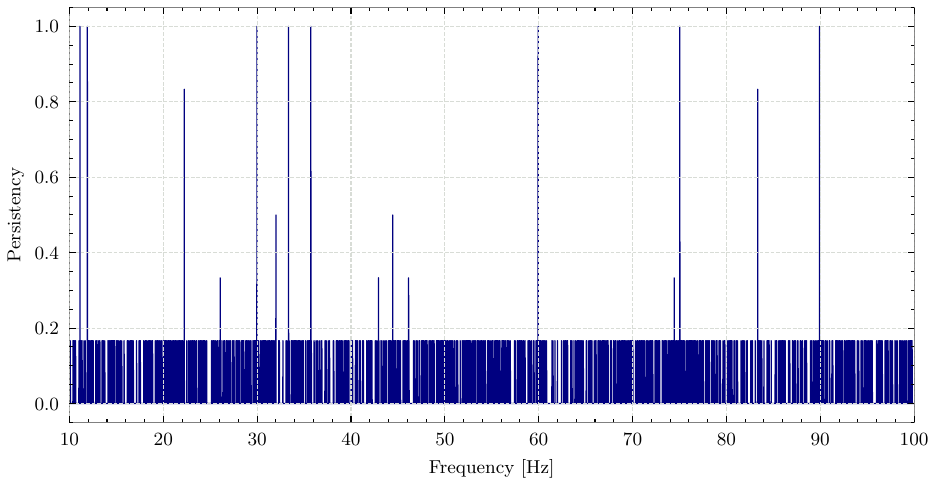}
    \caption{Example persistency spectrum produced by the low-noise, high-sensitivity $h(t)$ data from \ac{LIGO} Hanford detector on 1 Jan 2025 00:00 UTC to 2 Jan 2025 00:00 UTC. In this case, $N_{\rm e}=6$~segments.}
    \label{fig:fscan-persistency}
\end{figure}

The time-variation of a narrowband spectral artifact can sometimes provide crucial clues as to the nature of the cause of the artifact.
When an artifact becomes visible or vanishes at times that can be associated with detector configuration changes, it can help to localize the source of an artifact.
Persistency is computed using the $Z$-statistic
\begin{equation}
    Z_j(f_n) = \frac{P_j(f_n)-\mu_j(f_n)}{\sigma_j(f_n)}
\end{equation}
where $j$ indexes an interval of time that is longer than $T_{\rm SFT}$ and shorter than $T_{\rm avg}/2$ so that there will be more than one segment of time in a given $T_{\rm avg}$ interval, $P_j(f_n)$ is the average frequency bin power in the interval, $\mu_j(f_n)$ is the running mean and $\sigma_j(f_n)$ is the running standard deviation of $P_j(f_n)$ in interval $j$ running over $f_n$, respectively.
The running mean and standard deviation are typically 21~bins, $\approx$11.7~mHz.
The final persistency metric is computed as
\begin{equation}
    \mathcal{P}(f_n) = \frac{1}{N_{\rm e}}\sum_{j=0}^{N_{\rm e}-1}1(Z_j(f_n)\geq Z_{\rm thresh})
\end{equation}
where $1(\cdot)$ is the indicator function, $N_{\rm e}$ is the number of segments in a given averaging interval, and $Z_{\rm thresh}$ is a threshold to consider that the power in a frequency bin is above a certain amount, typically set to be 3.
An example persistency spectrum, $\mathcal{P}(f_n)$, is shown in \fref{fig:fscan-persistency}.

\subsubsection{\label{sec:linecount}Line counting}
Counting the number of narrowband spectral artifacts is important for understanding how corrupted the data may be from a high level.
In order to count lines in an automated fashion, a method is needed that is robust to a variety of spectral shapes and features while reliably counting lines.
Historically, the method assumes the normalized-averaged spectral data behaves with $\chi$-squared statistics, setting a certain threshold, and then counting the number of bins with normalized, average power higher than this threshold.
This method is simple, but the input data may not be $\chi$-squared distributed, the artifacts may be wider than single bins (leading to overcounting), etc.
A more robust method was employed for the O4 run, relying on methods designed for topographic prominence to count the number of peaks in a spectra~\citep{2020SciPy-NMeth}.
This leads to more robust line counts to overcome drawbacks of the thresholding method.

In this implementation, one must account for the fact that different averaging intervals may contain differing number of \acp{SFT}.
A spectrum containing only one \ac{SFT} will appear to contain many more prominences than a spectrum that has many \acp{SFT} because the spectral average with fewer \acp{SFT} has higher variance.
The number of prominences scales roughly like the inverse square-root of the number of \acp{SFT}.
Therefore, we have the option to normalize the number count of prominences by the square-root of the number of \acp{SFT}.
Although this is only a rough approximation, it is reasonable for the purposes of monitoring the data.

In addition to the total number of lines, we also compute the line count per frequency band, i.e., a line count density.
Frequency bands are chosen to characterize known detector issues.
As discussed in more detail in \sref{sec:case:violin-mode-contamination}, counting lines in specific frequency bands has proven useful to correlate with changes in detector performance.

In addition to employing improved line finding during daily, weekly, and monthly Fscan production epochs, we have developed a new scheme for finding combs among the lines found during each epoch.
This comb-finding scheme relies on nearest-neighbour pair-wise matching of lines and attempting to find common spacing of other found lines.
When a sufficient number of lines are grouped with common spacing, typically 5 or more, we then consider this to be a candidate comb that is saved for that epoch.
Observing these candidate combs over different epochs permits tracking of artifacts for further investigation (see \sref{sec:case-studies}).

\subsubsection{\label{sec:fscan-summary-pages}Fscan summary pages}
All of these Fscan monitors are assembled together into webpages available for on-site and off-site scientists within the \ac{LVK}, known as Summary Pages~\citep{Soni_2025}.
In O4, the Fscan webpages were more tightly integrated into the Summary Pages for better access and easier data monitoring by on-shift scientists monitoring data quality during the observing run.
A crucial step is creating a few simple figures-of-merit to enable at-a-glance assessment whether the daily, weekly, or monthly data shows significant contamination due to narrowband spectral artifacts.
This has proven invaluable to distribute the effort needed to carefully monitor the data for narrowband spectral artifacts.

\subsection{\label{sec:STAMP-PEM}STAMP-PEM} 
As discussed in \sref{sec:intro}, \ac{SGWB} searches are sensitive to both broadband and narrowband persistent spectra artifacts.
The cross-correlation analyses are carried out at comparatively coarser frequency resolution than \ac{CW} analyses.
Therefore, similar to the way higher frequency-resolution Fscan is important for \ac{CW} analyses, a lower frequency-resolution noise investigation method is important for \ac{SGWB} analyses.

We employ STAMP-PEM, a Python-based software package designed to compute coherence at coarse frequency-resolution between $h(t)$ and many auxiliary channels~\citep{Meyers:2018nyo}.
In contrast to Fscan, the coarse frequency-resolution of STAMP-PEM enables identification of broadband or moderately narrow spectral features in $h(t)$ that impact \ac{SGWB} analyses.
While running on low-noise, high-sensitivity data, STAMP-PEM calculates coherence  with auxiliary channels averaged on a daily timescale starting at 00:00:00 UTC. 
To manage data volume, STAMP-PEM computes \acp{SFT} using 10~s coherent length, then averages over longer timescales (typically 1800~s) before being saved.
Storing averaged quantities permits a reduction in the data volume in order to compute coherence over a larger number of auxiliary channels.
Tables are produced that list the highest coherent auxiliary channel and associated coherence value of every frequency bin; these tables are referred to as ``BruCo'' (brute force coherence) tables.

\subsection{\label{sec:Stochmon}StochMon and coherence outliers in cross-correlated spectra}
Both Fscan and STAMP-PEM produce coherence spectra between $h(t)$ and auxiliary channels at a given detector.
In principle both of these tools can also calculate coherence between any two channels, even between different detectors.
In the context of \ac{SGWB} searches, understanding the coherence between spatially separated detectors is important since the analysis relies on the assumption that detector noise is uncorrelated between detectors.
StochMon is a Python-based software package that mimics an isotropic \ac{SGWB} analysis to monitor $h(t)$ data from spatially separated detectors with the primary goal of identifying statistically significant coherence outliers violating this assumption~\citep{Stochmon}.

StochMon compares measured coherence values to the expected distribution under the null hypothesis of uncorrelated Gaussian noise.
For coherence estimates using Welch averaging, the relevant quantity is the effective number of independent segments, $N_{\mathrm{eff}}$, which accounts for correlations introduced by overlap and windowing. 
Under this null hypothesis, the upper-tail probability for observing a coherence value greater than $\gamma_0$ is
\begin{equation}
    P(\gamma > \gamma_0) = (1 - \gamma_0)^{N_{\mathrm{eff}} - 1} \, .
\end{equation}
This provides a $p$-value assignment for each frequency bin.
Typically, the frequency bins have width $1/32$~Hz.

To identify significant deviations, we adopt a false alarm rate (FAR) based approach. 
For a coherence spectrum with $N_f$ frequency bins, a threshold $\gamma_0$ is defined such that the expected number of bins exceeding this value under the null hypothesis equals a chosen FAR, giving
\begin{equation}
    \gamma_0 = 1 - \left(\frac{\mathrm{FAR}}{N_f}\right)^{1/(N_{\mathrm{eff}} - 1)} \, .
\end{equation}
Equivalently, frequency bins with
\begin{equation}
    p(f_n) = (1 - \gamma(f_n))^{N_{\mathrm{eff}} - 1}
\end{equation}
satisfying $p(f_n) < \mathrm{FAR}/N_f$ are identified as statistically significant coherence outliers.

Averaging intervals for StochMon are typically 1~day and 1~week, where daily averaging occurs over each day starting at 00:00:00 UTC and weekly averaging occurs over each week starting at 00:00:00 Monday UTC.
In addition to daily and weekly averaging, the cumulative results are produced on a daily basis in which the coherence is calculated using all data from the beginning of the corresponding observing epoch up to the given day.

\subsection{Line history studies}
As noted in \sref{sec:persistFOM}, recognizing changes to narrow spectral artifact behaviour and correlating those changes to detector configuration changes can provide evidence for an artifact source or coupling mechanism.
Historically, it has proven difficult to identify these change points due in part to limitations of person power with expertise in narrow line artifacts.
In O4, the line counting statistics (\sref{sec:linecount}) displayed as scatter plots on Fscan Summary Pages (\sref{sec:fscan-summary-pages}) have improved the identification of change points.
Understanding the history requires maintaining intermediate data products (e.g., normalized, averaged spectral data) over the whole observing run.

Aside from monitoring the line count history, two other investigations utilizing historical data have proven useful to identify change points in spectral artifact behaviour.
First, when investigating a particular artifact or set of artifacts, scatter plots showing power in those frequency bins over time can reveal these change points (see \sref{sec:case-studies}).
Second, using a hierarchical clustering algorithm derived from SciPy hierarchical clustering methods~\citep{scipy_18736568}, spectral artifacts at disparate frequencies with similar time varying behaviour can be associated.
Both of these investigation methods utilize data that is generated during daily, weekly, and monthly Fscan averaging.
This allows for discovering previously unknown relationships in spectral artifacts.

\subsection{Use of auxiliary channel data}
As described in \sref{sec:fscan-monitor} and \sref{sec:STAMP-PEM}, in addition to $h(t)$ data, many other auxiliary interferometric sensor data and physical environmental sensor data are stored.
These channels are broadly useful for maintaining low-noise, high-sensitivity detector operations and to monitor the local environment for non-astrophysical noise (e.g., stray magnetic fields, acoustic noise, and seismic motion).
Many \ac{DetChar} tools utilize these channels to monitor for transient noise artifacts and narrowband persistent noise artifact monitoring is no different.
We use Fscan and STAMP-PEM as complementary tools because Fscan can generate high-resolution coherence with auxiliary channels over long averaging time scales but over fewer channels; STAMP-PEM monitors more channels at lower frequency-resolution and shorter averaging time scales.

Within Fscan, these data channels are processed similarly to $h(t)$; \acp{SFT} are created, spectra and spectrograms are created.
In addition, the coherence (\sref{sec:coherence}) of the auxiliary channels and $h(t)$ is computed and plotted for display on the Fscan Summary Pages.
Evaluating the safety of auxiliary channels (i.e., some auxiliary channels may contain correlated information with the $h(t)$ channel due to control loop or environmental coupling) in the context of persistent spectral artifacts is a present topic of investigation.
Currently, we rely on the short duration transient tests to evaluate channel safety~\citep{Soni_2025}.
Additionally, the threshold for coherence in an auxiliary channel that triggers a concern for contamination is another topic of significant interest.
At present, we treat coherence with auxiliary channels as a contributing fact, but we do not rely solely on this information to determine the cause of a given narrow spectral artifact.

During periods of detector downtime, it has sometimes proven useful to continue to monitor auxiliary channels that are not dependent on a fully operational interferometer in order to monitor for configuration changes, primarily the physical environment monitor channels.
This data may be subject to other influences during site installation, commissioning, and maintenance activities, so careful evaluation is necessary.

Given the complementary strengths of different \ac{DetChar} tools, it is beneficial to compare the results, in this case between Fscan and STAMP-PEM.
Lines and combs identified in Fscan are compared against coherence results produced by STAMP-PEM.
Frequencies of interest from the higher frequency-resolution results from Fscan are scaled to the corresponding lower frequency-resolution STAMP-PEM results.
Frequencies of interest are identified in BruCo tables to identify particular auxiliary channels that may be witnessing the artifact.
Over multiple days, if the same channel is identified multiple times for the same frequency bin, further investigation studies are carried out.

In addition to the Fscan and STAMP-PEM programs, other tools leverage their outputs in meta-analyses.
One such example is a ``Linefinder'' tool that provides a way to navigate through different auxiliary channel outputs of Fscan and STAMP-PEM programs~\citep{Coughlin_2010}.
Linefinder organizes their spectra and coherence results into an alternative searchable format that can filter by date, frequency, and channel.
This allows identifying potential auxiliary channels of interest when investigating specific line frequencies across longer durations.
It is also useful when there are changes in coherence or spectral features indicating possible changes in the noise source or coupling.

\subsection{Prioritization of investigation efforts}
Due to the multitude of lines observed in both detectors during previous Advanced \ac{LIGO} observing runs and our limited resources to investigate each of them, we face the challenge of determining which narrowband artifacts should be prioritized over others.
We have developed an astrophysically-motivated procedure which takes into account: how prolific, i.e., the fraction of the total band the artifact impacts, whether the artifact is impacting data in a more sensitive band of the detector, whether the artifact may be impacting data from specific known pulsars, and extending the known pulsar population as representative for the galactic population of unseen neutron stars.
Taking these factors together, we calculate a priority rating score for each artifact; note that a comb is considered one artifact that impacts multiple frequency bins.
This procedure allows us to allocate limited resources to focus on those artifacts which, if mitigated, would benefit narrowband persistent \ac{GW} searches.
During O4, we prioritized efforts on several comb artifacts which we discuss in \sref{sec:case-studies}.
In the future, we plan to integrate this prioritization scheme into production Fscan in order to better monitor and prioritize work throughout observing epochs.
A detailed description of this procedure will be presented in a forthcoming publication (Starkman, in prep.).

\section{\label{sec:case-studies}Case studies}
In this section, we describe a number of examples of particular narrow spectral artifacts from the two \ac{LIGO} detectors data from O4 that were investigated and, where possible, mitigated.
The O4 run is divided into three epochs: O4a (24 May 2023 15:00 UTC to 15 Jan 2024 16:00 UTC), O4b (10 Apr 2024 15:00 UTC to 28 Jan 2025 17:00 UTC), and O4c (28 Jan 2025 17:00 UTC to 18 Nov 2025 16:00 UTC).
Other case studies from the first two Advanced \ac{LIGO} observing runs (O1 and O2) are described in~\citep{PhysRevD.97.082002}.
During O4, we focused almost all of our investigation efforts to the \ac{LIGO} Hanford detector because there are many more narrowband persistent artifacts observed in the Hanford data than the Livingston data.

\subsection{OM2 heater driver comb}
\begin{figure}
    \centering
    \includegraphics[width=1\textwidth]{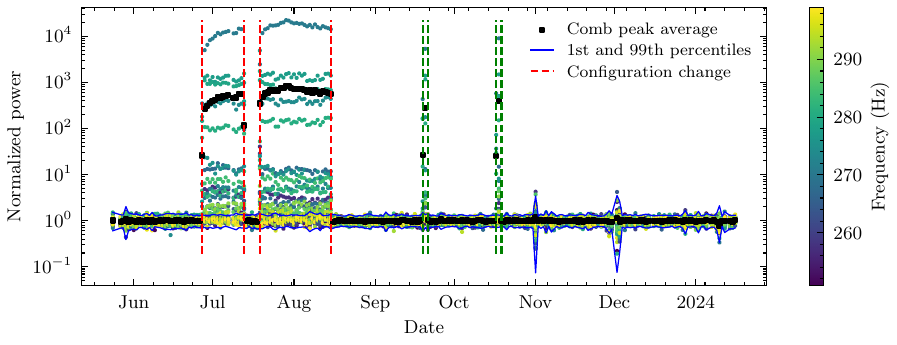}
    \caption{Scatter plot showing the normalized power for the frequency bins impacted by the $\approx$1.6611~Hz comb for each day during O4a. Colored scatter points indicate different frequency bins, black points are the peak average, the blue curves are the 1st and 99th percentiles of the frequency band selected (250-300~Hz), and dashed lines show configuration changes. In dashed red are the OM2 optic heater turning on or off. Note that the comb increases significantly when the heater is turned on. In dashed green are efforts to change the electrical connections to mitigate the comb, where two of the efforts were unsuccessful.}
    \label{fig:1.66Hz}
\end{figure}

Prior to the start of and into early O4, a comb artifact with spacing of $\approx$1.6611~Hz was observed in \ac{LIGO} Hanford $h(t)$ spectral data, with over 200 harmonics visible~\citep{alog:66418,alog:66925}. 
With so many harmonics this comb was an early target of our investigations.
The comb was not found in \ac{LIGO} Livingston spectral data.
During our investigations, we found that the comb disappeared and reappeared at a few dates during this early O4 period, see \fref{fig:1.66Hz}~\citep{alog:69791,alog:71108}.
There were no recorded logs of interferometer configuration changes made during those dates where the comb behaviour changed. 
Later, a summary of configuration changes made to a particular optical heating element (the output mirror 2, OM2, optic) from the previous several months was reported~\citep{alog:71533}. 
Shortly thereafter, we established the relationship where changing the OM2 heater setpoint caused the comb to appear or disappear~\citep{alog:71801}.

Detector scientists and engineers changed the electrical connections to the Beckhoff controller element for the OM2 heater, and with this change, the comb was no longer visible in the daily Fscan spectra~\citep{alog:72269}.
Early efforts at more permanent solutions caused the comb to return briefly as shown in \fref{fig:1.66Hz}, but an improved configuration change successfully mitigated this comb.

\subsection{HWS camera shutter comb}
\begin{figure}
    \centering
    \includegraphics[width=1\textwidth]{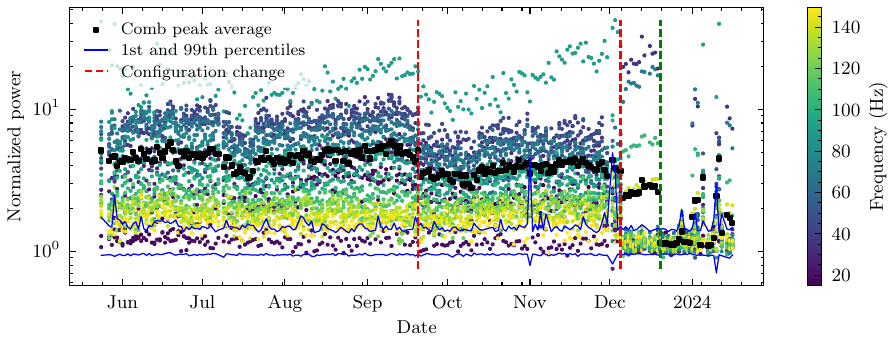}
    \caption{Scatter plot showing the normalized power for the frequency bins impacted by the $\approx$4.9842~Hz comb for each day during O4a. Colored scatter points indicate different frequency bins, black points are the peak average, the blue curves are the 1st and 99th percentiles of the frequency band selected (14-150~Hz), and dashed lines show configuration changes. In dashed red are the changes to the HWS software control. Note that the comb behaviour changes at the same time. In dashed green are the successful change in detector operations to mitigate this comb.}
    \label{fig:4.98Hz}
\end{figure}

Another comb, first observed in LIGO Hanford $h(t)$ data before the start of O4, with a spacing of $\approx$4.9842~Hz impacted O4 data with over 100 harmonics visible, see \fref{fig:4.98Hz}~\citep{alog:66925,alog:68261}.
At other times during early O4, other combs with spacing $\approx$0.9968~Hz~\citep{alog:74585} and $\approx$6.9779~Hz~\citep{alog:74617} were observed at various points, also with large number of harmonics visible.
The $\approx$0.9968~Hz comb had also been observed in O3 data, but the source was not identified at the time~\citep{T2100200}.
During O4a, these near-integer frequencies were at last identified by site engineers to be related to the input test mass Hartmann Wavefront Sensor (HWS) camera shutter rates that had been employed during normal operations~\citep{alog:74750}.
Once changes were made to the way the HWS camera operates during the detector low-noise, high-sensitivity running, the combs were effectively mitigated~\citep{alog:74900}.

\subsection{PSL flow sensor combs}
\begin{figure}
    \centering
    \begin{subfigure}{1\textwidth}
        \includegraphics[width=1\linewidth]{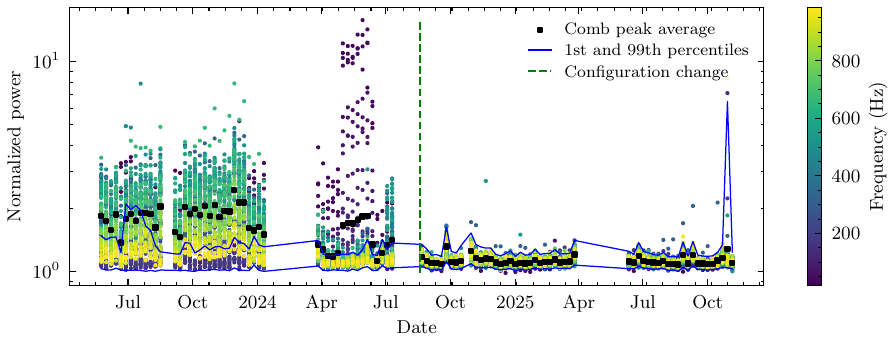}
    \end{subfigure}
    \begin{subfigure}{1\textwidth}
        \includegraphics[width=1\linewidth]{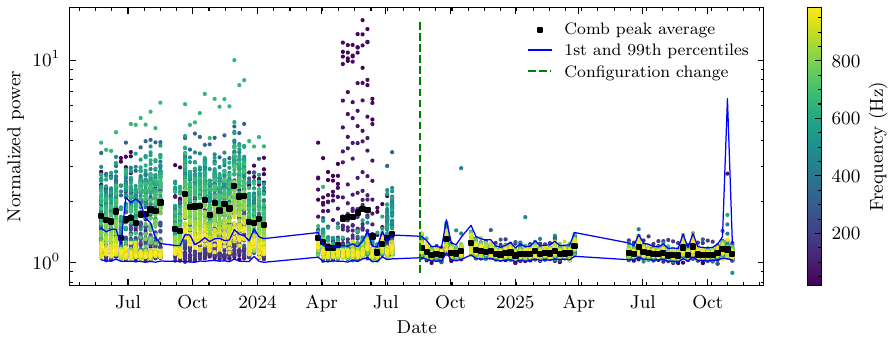}
    \end{subfigure}
    \begin{subfigure}{1\textwidth}
        \includegraphics[width=1\linewidth]{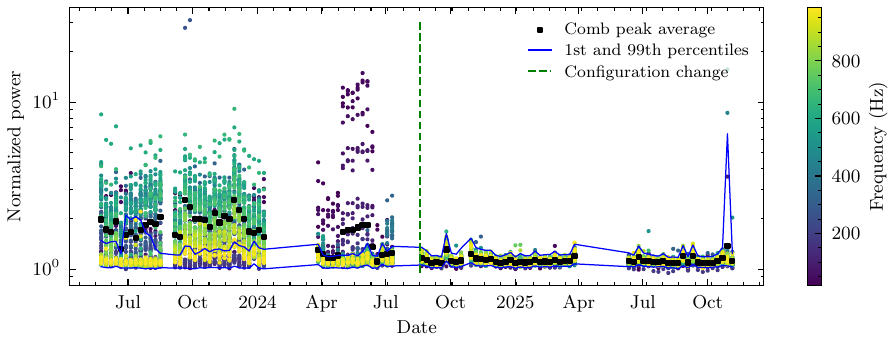}
    \end{subfigure}
    \caption{Scatter plots showing the normalized power for the frequency bins impacted by the $\approx$9.48~Hz combs (upper: $\approx$9.474~Hz, middle: $\approx$9.475~Hz; lower: $\approx$9.48~Hz) for each week during O4 \ac{LIGO} Hanford data. Colored scatter points indicate different frequency bins, black points are the peak average, the blue curves are the 1st and 99th percentiles of the frequency band selected (18-994~Hz), and the dashed green line shows the configuration change to mitigate this comb.}
    \label{fig:9.48combs}
\end{figure}

Throughout much of O4a and O4b, a series of combs with artifacts spaced near $\approx$9.48~Hz with over 100 harmonics each was observed in LIGO Hanford and only a few harmonics observed in LIGO Livingston $h(t)$ spectra~\citep{alog:68261,alog:70219,alog:70030}.
This comb fluctuated slightly in frequency and had changes in amplitude behaviour.
These changes, unfortunately, did not align well with specific configuration changes such that we could identify a cause for this comb; the behaviour often changed during longer breaks in operations when many configuration changes were made, see \fref{fig:9.48combs}.
Since these artifacts were not observed in LIGO Hanford O3 data, we suspected that if we could determine when the comb first appeared in between runs, then we could more easily identify the source.
In order to track this comb artifact in time, however, we needed to first identify a witness channel that did not rely on the detector being in low-noise, high sensitivity operation.

Using the combination of Fscan coherence and STAMP-PEM coherence, we identified a subset of channels from the Pre-Stabilized Laser (PSL) subsystem that could be witnessing this noise~\citep{alog:70219}.
Manual inspection confirmed a small number of channels that could be used in place of $h(t)$~\citep{alog:77990}.
There were several initial checks of the PSL subsystems at both LIGO sites to ascertain the cause of this noise, but nothing conclusive was determined from these checks.
Manual inspection of LIGO Hanford historical data spectra found that the comb appears in the witness channels on or near 2 September 2021, around the time major PSL upgrade work was ongoing~\citep{alog:78142}.
With the insight that something around the PSL must be causing the noise, a portable magnetometer was employed while a controlled PSL shutdown was initiated to determine the cause~\citep{alog:79533}.
Finally, the source was determined to be PSL chiller flow meter sensors installed as part of the PSL upgrade work prior to the start of O4.
Site engineers reworked the power supply for the PSL chiller flow meter sensors in order to isolate this noise, which mitigated the noise in $h(t)$~\citep{alog:79593}.
This fix was also employed at LIGO Livingston~\citep{alog:73376}.

\subsection{\label{sec:case:violin-mode-contamination}Violin mode contamination}

\begin{figure}
    \centering
    \includegraphics[width=1\textwidth]{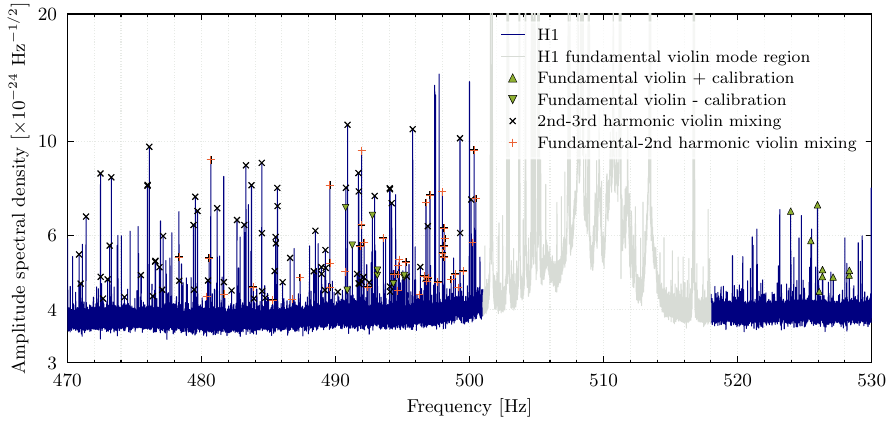}
    \caption{\ac{LIGO} Hanford $h(t)$ amplitude spectral density (blue curve) in the 470~Hz to 530~Hz frequency band during the week of 28 June 2023 to 4 July 2023 showing severe line contamination~\citep{alog:82320}. The frequency band containing the fundamental violin mode is grayed out to draw focus to the abnormally contaminated bands around the normally contaminated band. In the abnormally contaminated bands (blue curve), line artifacts are marked according to different intermodulation sources: violin mode frequencies and calibration lines (upward triangles, additive; downward triangles, difference), second and third violin mode harmonics (black crosses), and fundamental and second harmonics (orange pluses). Similar line artifacts caused by intermodulation are observed around the second and third violin modes.}
    \label{fig:violin_mode_contamination}
\end{figure}

The quadruple suspension systems of the Advanced LIGO detector test masses~\citep{Aasi2015} exhibit
well known violin modes, corresponding to mechanical resonances of the suspension fibers. 
The fundamental modes are observed near $\sim$500~Hz, with higher-order harmonics near $\sim$1000~Hz and above~\citep{ViolinMode} (see \fref{fig:O4spectra}). 
These narrowband artifacts are present continuously in the data. 
At various times during O4, the violin modes were excited due to motion of the suspension points, raising the amplitude of these persistent artifacts. 
When that happened, we also observed an appearance of lines in the surrounding frequency bands at \ac{LIGO} Hanford normally clean of artifacts~\citep{alog:71501,alog:71800}.
These additional lines would be present for hours to days until the violin modes were damped. 
Although the additional lines are not present for the entirety of O4, they nevertheless are observed in the run-average spectra and degrade persistent \ac{GW} searches, as shown in \fref{fig:violin_mode_contamination}.

Since this additional contamination is only observed at \ac{LIGO} Hanford, we performed a comparison between Hanford and Livingston to check whether differences in the detector configuration could explain these artifacts.
Investigations are challenging because we do not wish to purposefully excite violin resonances since they are hard to damp once excited.
We therefore must examine historical data or wait until the next violin mode excitation.
In addition, not all violin modes are excited equally, and we are not sure if there is one problematic mode or many.

\begin{figure}
    \centering
    \includegraphics[width=1\textwidth]{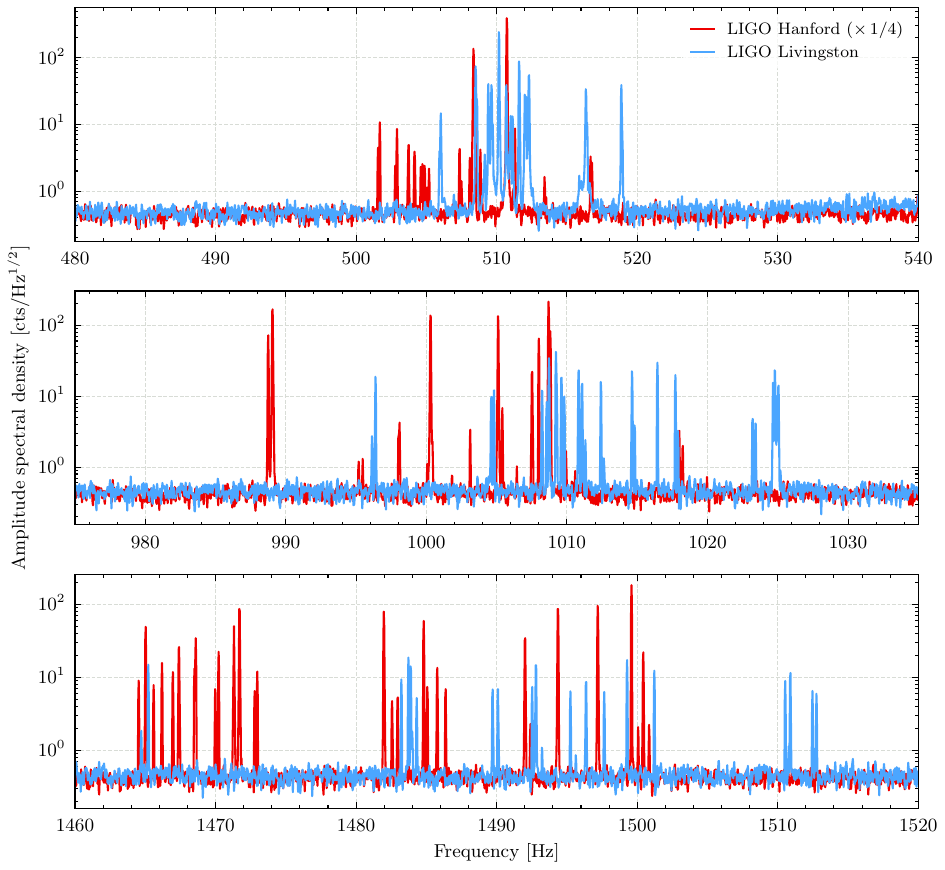}
    \caption{Comparison of analog-to-digital converter (ADC) spectra between \ac{LIGO} Hanford (red curve) and \ac{LIGO} Livingston detector (blue curve) in the frequency bands containing the violin modes. Upper panel: violin first harmonic (fundamental) modes; middle panel: second harmonic modes; lower panel: third harmonic modes. The LIGO Hanford ADC counts are scaled by a factor of $1/4$ to account for the fact that they are recorded as the sum of four photodiodes.}
    \label{fig:violin_ADC_count_comparison}
\end{figure}

While examining historical data, we found that the effective photon counts in the interferometric photodetector readout at higher order violin harmonics (e.g., $\sim$1000 and $\sim$1500~Hz) are significantly larger at \ac{LIGO} Hanford than at Livingston~\citep{alog:84136}, see \fref{fig:violin_ADC_count_comparison}.
Additionally, we observed that when the amplitude of the second violin mode is higher, the number of narrow spectral lines in the surrounding region of the fundamental mode ($\sim$500~Hz) also increases~\citep{alog:84136}.
This behaviour was observed similarly across different O4 epochs.

The relationship between violin-mode amplitude and excess line contamination, however, appeared to be epoch dependent and not set by a single amplitude threshold for all epochs~\citep{alog:71501,alog:82320}.
Further evidence indicated that the excess line contamination is not determined by violin-mode amplitude alone, but may also depend on the presence of additional spectral lines, such as calibration lines~\citep{alog:82320}.
This motivated investigation of the physical origin of the excess narrow-band structure.

Subsequent line-identification studies suggested that many of these narrow spectral artifacts are consistent with intermodulation arising from non-linear effects in the interferometric readout electronics.
We suspected intermodulation between violin modes and purposefully added calibration lines at low frequencies~\citep{alog:87923}.
If there was a small non-linear term of the interferometric readout electronics that can be approximated by a linear plus a small quadratic term, then we could predict the intermodulation frequencies of calibration line and violin modes and between different violin modes~\citep{alog:87923}.
This non-linear term would be most evident by the loud violin modes and the loud, low frequency calibration lines.
The predicted intermodulation frequencies were compared to the frequencies of the excess line contamination~\citep{alog:87923}.
This comparison suggested that many of the excess narrow spectral lines are consistent with frequencies predicted by a simple quadratic intermodulation model involving violin modes, calibration lines, and their harmonics, see \fref{fig:violin_mode_contamination}.

We then estimated the proposed non-linear, quadratic term for each of the intermodulation frequency pairings~\citep{alog:87923}.
The quadratic term was estimated at two points along the interferometric readout; from analog (milliamps of photocurrent) and after the analog-to-digital converter (digital counts)~\citep{alog:87923}.
This comparison helps to constrain where a potential non-linear term may be introduced, i.e., whether the observed intermodulation is consistent with non-linear effects in the photodetector electronics or digital-to-analog conversion.
These estimates indicated that a qualitatively similar intermodulation patterns exists in both domains but were inconclusive in providing a localization of where along the readout the effective quadratic non-linearity is introduced.

A conclusive explanation has not yet been reached for different violin mode behaviour between \ac{LIGO} Hanford and Livingston detectors.
While both detectors employ some degree of active violin mode damping, only \ac{LIGO} Livingston currently employs passive violin mode damping hardware.
At present, it remains unclear the cause for challenging violin mode performance at \ac{LIGO} Hanford.
Further exploration and characterization of damping techniques are under development.

\subsection{\label{sec:case:calline-intermod}Temporary calibration line intermodulation}
In O4, the \ac{LIGO} calibration team implemented new, continuously running calibration monitoring lines at both sites (at slightly different frequencies) in addition to previously running calibration lines.
Early in O4, a number of temporary calibration lines were added at \ac{LIGO} Hanford via the photon calibrator subsystem~\citep{10.1063/1.4967303} and through the interferometric servo control loop that controls the differential-arm degree-of-freedom~\citep{alog:71706}.
These lines on their own are identified and recorded through detector operations; their frequency and start and stop times are known.
Calibration line intermodulation between other calibration lines or suspension violin modes has been observed in previous observing runs~\citep{T2100200}.

\begin{figure}
    \centering
    \includegraphics[width=1\textwidth]{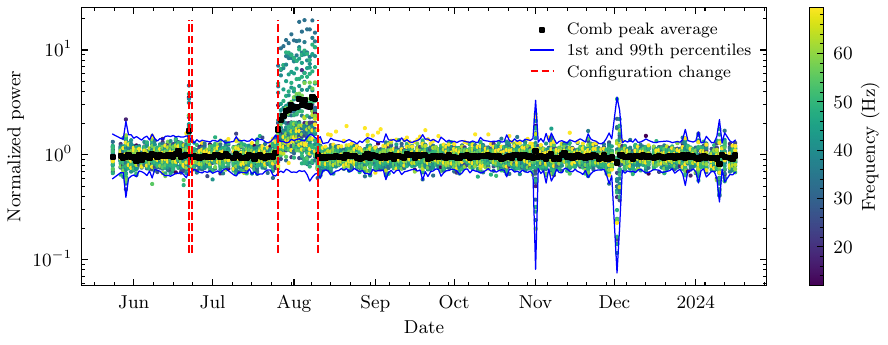}
    \caption{Scatter plots showing the normalized power for the frequency bins impacted by the temporary calibration line intermodulation between different calibration lines in \ac{LIGO} Hanford data. Colored scatter points indicate different frequency bins, black points are the peak average, the blue curves are the 1st and 99th percentiles of the frequency band selected (10-70~Hz), and the dashed red lines show when the lines were turned on or off.}
    \label{fig:calline-intermod}
\end{figure}

It was not known, however, the extent to which a small number of additional lines would exacerbate line contamination issues, especially when suspension violin modes are excited beyond their nominal levels.
A large number of additional lines are observed when the violin modes are excited, but they are difficult to associate without detailed study~\citep{alog:79825}.
Intermodulation between different calibration lines (temporary and permanent) during O4a is shown in \fref{fig:calline-intermod}, even during periods of lower violin mode amplitudes.
Once the connection of the large number of additional lines and the temporary calibration lines was established, these temporary lines were turned off~\citep{alog:71964}.

At present, intermodulation between different calibration lines and calibration lines with violin suspension resonances remains problematic.
Additional calibration lines may be an attractive way to monitor the length-response of detectors, but, as we have seen, their use can detrimentally impact detection of narrowband persistent \ac{GW} signals.

\subsection{Aliasing artifacts}
\label{sec:duotone}
\begin{figure}
    \centering
    \includegraphics[width=1\textwidth]{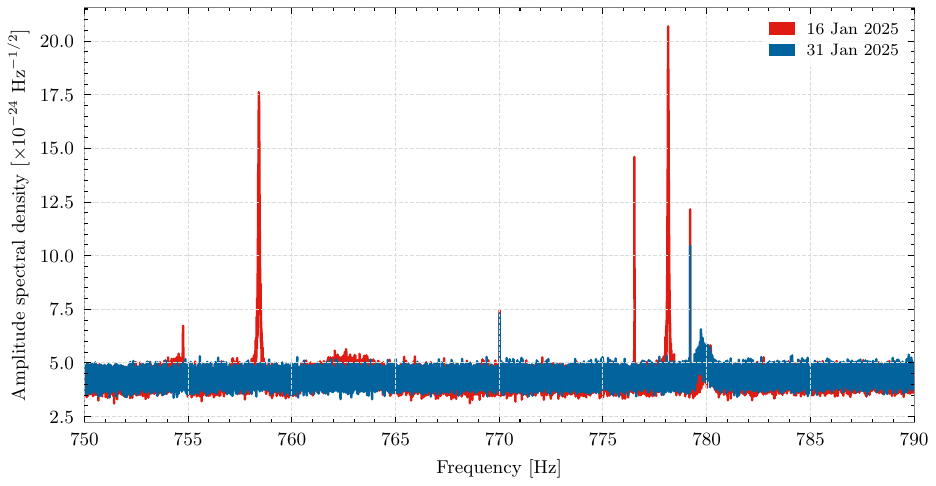}
    \caption{Daily spectra of low-noise, high-sensitivity $h(t)$ from 16 Jan 2025 (red curve) and 31 Jan 2025 (blue curve) showing an example of aliasing lines in the 750-790~Hz band, before and after improved anti-aliasing filtering was installed, respectively.}
    \label{fig:aa-artifacts}
\end{figure}

As part of investigations related to the causes of violin mode contamination (\sref{sec:case:violin-mode-contamination}) and temporary calibration line intermodulation (\sref{sec:case:calline-intermod}), we ran diagnostics on the \ac{LIGO} Hanford digital path of the interferometric readout photodetectors.
These diagnostic checks revealed aliasing artifacts in-band due to insufficient anti-alias filtering~\citep{alog:82329}.
Once additional filtering was installed~\citep{alog:82375} several narrowband persistent artifacts above 400~Hz were mitigated~\citep{alog:82455,alog:82512}, see \fref{fig:aa-artifacts}.

\subsection{Near-30 Hz and near-100 Hz comb families}

Throughout O4, two sets of prolific combs were observed in the \ac{LIGO} Hanford $h(t)$ data but not observed in Livingston $h(t)$ data: a near-30~Hz comb ($\approx$29.9695~Hz) and near-100~Hz comb ($\approx$99.998~Hz).
Besides the comb occurring at frequencies associated with harmonics of these frequencies, a number of other combs with the same spacing but different frequency offsets were observed, leading to suspicion of intermodulation with violin resonances as well as each other.
The combs also did not appear as single peaks in the run-averaged data, but rather multiple peaks at closely spaced frequencies.
This could be due to several sources of the comb or indicate a drift of the frequency generating the comb during O4.
We refer to these various different combs as the near-30~Hz family and near-100~Hz family.
Several of these combs were also found in O3 \ac{LIGO} Hanford data~\citep{T2100200}.

Investigations using auxiliary channels indicated that the near-30 Hz comb family originated within the main building housing the main interferometer laser, input mode cleaner, input mirrors and beamsplitter, and readout and electronics, whereas the near-100 Hz comb family were observed at both the main building and end buildings containing end mirrors.
This difference suggests two different sources or coupling mechanisms~\citep{alog:87414}.
Several early investigations on whether there was a correlation with input mirror electrostatic drive bias settings or grounding changes provided relatively modest insights~\citep{alog:87889,alog:88089}.

Using a portable magnetometer, a measurement campaign localized the $\approx$29.9695~Hz fundamental frequency to the Pre-Stabilized Laser (PSL) Interferometer Sensing and Control electronics rack in the main building. 
Further investigation of the rack identified a 12~V distribution and fused power supply feeding three video cameras in the PSL enclosure.
When disconnecting the 12~V distribution or removing fuses, the fundamental peak disappeared, and reconnecting caused the fundamental peak to reappear.
The near-100~Hz combs, however, were unaffected by this change, providing confirmation that the sources are unrelated~\citep{alog:88595}.

\begin{figure}
    \centering
    \includegraphics[width=1\textwidth]{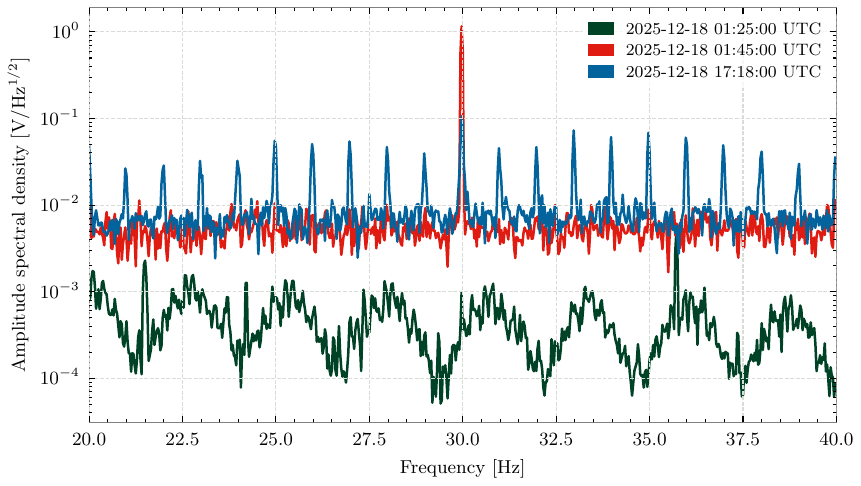}
    \caption{Comparison of amplitude spectral density (ASD) measured by a voltage monitor of the power supply for the three PSL Axis 215 PTZ cameras (\texttt{H1:PEM-CS\_ADC\_5\_23\_2K\_OUT\_DQ} channel) in the 20–40 Hz band at three different times on 18 Dec 2025. When all cameras were powered off, no visible 29.97~Hz peak was visible (green trace, 01:25 UTC). When the cameras were turned on, the 29.97~Hz fundamental frequency is evident (red trace, 01:45 UTC). As a test, the frame rate was changed on the three cameras from 30~frames/s to 1~frame/s and the voltage monitor ASD changed to show a 1~Hz comb with a reduced 29.97~Hz peak (blue trace, 17:18 UTC).}
    \label{fig:near30Hz}
\end{figure}

The three PSL video cameras (Axis 215 PTZ cameras) are NTSC-timed devices operating at a nominal 30~frames/s~\citep{axis215ptz_datasheet}, which, following the National Television System Committee (NTSC) standard, corresponds to an actual frame rate of 29.97~frames/s~\citep{NTCSstandard}.
While changing the video camera settings had small effects on the near-30~Hz comb, the only effective mitigation was powering off the video cameras (see \fref{fig:near30Hz}).
These cameras are not in operation at \ac{LIGO} Livingston, which is the reason that the Livingston data did not contain these combs.
Although we successfully identified the source of a problem that has persisted for many years, it was, unfortunately, not identified until the end of O4~\citep{alog:88595}.
These cameras are not required for low-noise, high-sensitivity operations, but were nonetheless left on during observing epochs.

\subsection{Suspected Beckhoff comb}
Combs with spacing near $\approx$11.904~Hz were observed in both \ac{LIGO} Hanford and \ac{LIGO} Livingston $h(t)$ data throughout O4, with the strongest visible harmonics at $\approx$11.904~Hz and $\approx$35.712~Hz, the odd harmonics of the comb~\citep{alog:66036,alog:72866}.
The comb has slightly different spacing in the two \ac{LIGO} detectors, and no significant coherence between \ac{LIGO} detectors was found at the odd harmonics, suggesting that the source was not synchronized to \ac{GPS} clocks.
On the other hand, at both detectors, coherence was observed between $h(t)$ and magnetometers located across the site buildings, suggesting some intra-site synchronization~\citep{alog:66036}.

Studies performed at \ac{LIGO} Livingston suggested the comb is caused by the Beckhoff system~\citep{alog:66385} and the coupling mechanism through the test mass electrostatic driver actuation electronics~\citep{alog:67581}.
A ground path coupling may be possible since the fundamental frequency is visible as a peak in a \ac{PSD} of the Beckhoff power supply.
Since the comb is witnessed by magnetometers located across the site buildings and the Beckhoff system is broadly deployed to the buildings as well, this suggests a likely culprit for the $\approx$11.904~Hz comb.
At present, this comb has not yet been mitigated at either \ac{LIGO} site.

\subsection{Intersite coherence study}
\label{sec:case:intersite_coherence}
\begin{figure}
    \centering
    \includegraphics[width=\textwidth]{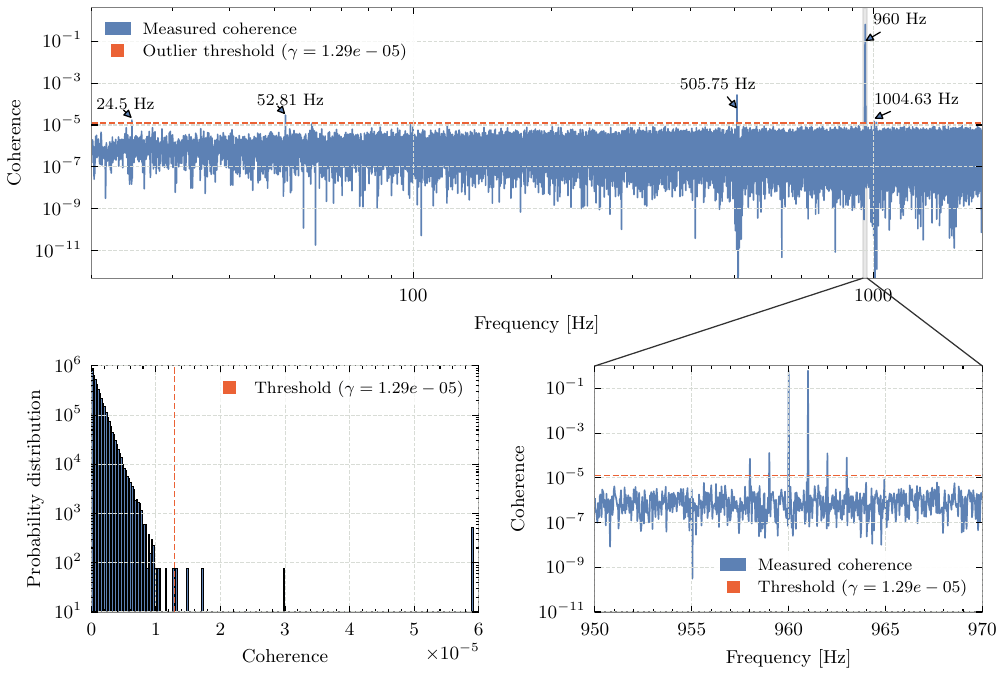}
    \caption{Coherence spectrum and coherence distribution observed during O4. Upper panel: coherence between \ac{LIGO} Hanford and \ac{LIGO} Livingston over the 20-1750~Hz band with frequency resolution $1/32$~Hz computed using the full O4 dataset. The horizontal dashed line indicates the coherence threshold expected for uncorrelated Gaussian noise. Annotated frequency bins above this threshold correspond to coherent narrowband artifacts. Lower left: distribution of coherence values with outliers forming the high-coherence tail to the right of the threshold marked with a dashed line. Lower right: zoom around the DuoTone artifacts, which are present at both detectors and exhibit the largest coherence above the threshold.}
    \label{fig:stochmon_coherence}
\end{figure}

We employed StochMon (\sref{sec:Stochmon}) to measure the coherence between $h(t)$ of the two \ac{LIGO} detectors during O4.
The upper panel of \fref{fig:stochmon_coherence} shows the coherence spectrum at a frequency resolution of $1/32$~Hz.
A threshold is calculated using a fixed FAR that corresponds to an expected 1 outlier per coherence spectrum under the null hypothesis.
Frequency bins with coherence value that exceeded this threshold were flagged for additional investigation.
For clarity, only the most prominent outlier features are annotated. 
The bottom left panel of \fref{fig:stochmon_coherence} shows the distribution of coherence values.
The bulk of the distribution is consistent with the expectation of uncorrelated noise behaviour while the outliers lie outside the bulk of the distribution.

The annotated outliers correspond to known instrumental features present in both detectors.
These include a temporary calibration line at 24.5~Hz that was active during a brief period of O4a~\citep{alog:72096} (\sref{sec:case:calline-intermod}), a simulated \ac{CW} hardware injection at 52.81~Hz, and harmonics of the suspension violin modes near 505.75~Hz and 1004.63~Hz. 
The most significant feature near 960~Hz is associated with the DuoTone component of interferometer timing system~\citep{PhysRevD.108.022003,alog:64525}.
Although these frequencies have been used since the initial \ac{LIGO} era in the 2000s, they have not typically been present in the $h(t)$ data historically.
Due to a redesign of the interferometric readout electronics for O4, however, these lines were introduced to both \ac{LIGO} Hanford and \ac{LIGO} Livingston $h(t)$ data.
The DuoTone lines are synchronized to \ac{GPS} clocks at both detectors and are thus highly coherent with one another.
As shown in the zoomed panel, this feature consists of multiple closely spaced lines: the primary frequencies, 960 and 961~Hz, as well as a 1~Hz comb as sidebands on the primary frequencies~\citep{alog:77696}.

A proposed electronics configuration change to mitigate the DuoTone artifacts was deemed inadequate for the timing system diagnostics, so the preferred solution was to move the DuoTone frequencies to 1920 and 1921~Hz (1~Hz separation, increasing the 960~Hz line by a factor of 2)~\citep{alog:82647,alog:72103}.
Although the frequencies at the two sites remains the same, leading to highly coherent artifacts at $\sim$1920~Hz, the frequencies were deemed high enough---at least for now---that they do not detrimentally impact narrowband persistent \ac{GW} searches to the same degree.

\section{\label{sec:dq-products}Data quality products}
In addition to investigating line artifacts, the \ac{LIGO} \ac{DetChar} group is responsible for producing data quality products to aid narrowband persistent \ac{GW} searches~\citep{Soni_2025}.
Specifically, two products are produced: ``lines lists'', primarily used by \ac{CW} searches, and ``notch lists'', primarily used by \ac{SGWB} searches.
These data quality products provide crucial information to searches in order to improve search sensitivity and avoid costly follow-up both in personnel and computational resources.

\subsection{\label{sec:line-lists}Lines lists}

The curation of the lines lists is a multi-stage procedure that makes use of both automated line-finding tools and visual inspection.
The initial stage involves automated identification of all peaks seen in normalized run-averaged spectra for each detector, where the spectra are generated from $T_\mathrm{SFT}=7200$~s self-gated $h(t)$~\citep{T2400003} \acp{SFT} averaged over some epoch of O4, as discussed in \sref{sec:spectral_averages}.
The peak-finding algorithm looks for topographic prominences in the spectra and tags them accordingly~\citep{goetz_2025_18776397, 2020SciPy-NMeth}, building a preliminary ``working list'' of line artifacts.
This is followed by visual inspection of the run-averaged spectra from $10$ to $2000$~Hz in order to verify that the automated tool has properly tagged all peaks in the spectrum; spurious, off-center, or missing tags are corrected by hand, where necessary.

All tagged peaks are then tabulated into either an ``unvetted'' list or a ``vetted'' list for the corresponding detector.
An artifact is added to the vetted list if it has been established to be non-astrophysical in origin, a conservative choice intended to maximize the amount of usable data and minimize the probability of false dismissals.
In order to establish that a specific narrow line or comb is non-astrophysical, we require that a known source must be identified even if the coupling mechanism is unknown; the source is established when the source is active and the line or comb is observed and when the source is inactive and the line or comb is no longer observed.
Alternatively, for combs only, the mere observation of multiple comb peaks ($>$5~peaks) is enough to establish the non-astrophysical nature of the artifact, since no emission mechanism is known to cause a \ac{NS} to emit at $>$5 frequency values\footnote{It is worth noting that single lines with very loud amplitudes, or that appear in only one interferometer, are not automatically added to the vetted list. Since we want to maximize the amount of usable frequencies for \ac{CW} searches, such artifacts are kept in the unvetted list until an instrumental identification is made. Also, in some cases where fewer than 5 comb teeth are visible, if the spacing is unlikely to be an astrophysical source, e.g. integer frequency spacing, then the comb is considered vetted.}.
Only artifacts in the vetted list are recommended for use as outlier vetoes, whereas the unvetted list is generally reserved for deeper data quality studies when an outlier cannot be ruled out via other methods.

For each spectral artifact, the vetted lists record the artifact type, central frequency (if a line) or frequency spacing (if a comb), left and right frequency widths, and, new for O4 and where available, \ac{GPS}-second time segments indicating when the artifact was known to be present. 
Additionally, for combs only, we record the frequency offset and number of visible harmonics.
In O4, we grouped spectral artifacts into four possible types: single lines, fixed width combs, scaling width combs, and a fourth type (``band'' artifact) that marks an entire band as heavily contaminated with multiple artifacts, which was specifically created to address the violin mode contamination  at H1.
Fixed width combs have constant frequency bandwidths across all comb teeth, whereas for scaling width combs the bandwidth of each tooth scales linearly with its harmonic number. 
The $60$~Hz power mains and its higher harmonics form one such example of a scaling width comb. 

Lines list production in O4 was a cumulative process.
We produced an initial version of the lines lists based on O4a run-averaged spectra, identifying the majority of line and comb artifacts present in the data~\citep{T2400204}.
As more data are accumulated, our run-averaged spectra expose quieter artifacts by virtue of having better statistics, in addition to revealing new artifacts introduced by mid-run detector configuration changes.
In total, we produced four versions of the lines lists covering increasingly longer epochs of O4: O4 start to end of O4a, O4 start to the end of O4b, O4 start to the end of the first half of O4c, and finally a complete version spanning all of O4.
Lines lists for a given epoch are occasionally updated when new information is available, for example, to add artifacts that were missed in the initial round of spectral inspections.

At the conclusion of O4, the total number of vetted artifacts (counting individual comb teeth) in each interferometer is as follows: 2399 in LIGO Hanford, with total vetoed bandwidth equal to $17.21\%$ of the $10$-$2000$~Hz observing band, and 449 in LIGO Livingston, equal to $6.37\%$ of the observing band.
A significant fraction of the vetoed bandwidth at Hanford is due to the excessive contamination around the violin modes. 
If such band artifacts are not counted, then the vetoed bandwidth at Hanford reduces to $5.92\%$\footnote{Although the \ac{LIGO} Hanford veto fraction excluding the violin mode contamination regions is smaller than \ac{LIGO} Livingston, there are more individual artifacts in frequency in the \ac{LIGO} Hanford data.}. 

\subsection{\label{sec:Notch_lists}Notch lists}
Notch lists are different than lines lists because it is a single data quality product that includes information from two (or more) detectors.
It contains information optimized for the statistical properties and sensitivity of cross-correlation analyses used in \ac{SGWB} searches. 
The lists target both narrowband and broadband spectral artifacts that can bias cross-correlation measurements.
Although there may be a common set of frequency bins noted in each data quality product, there are several differences.
First, notch lists are comparatively coarse in frequency (1/32~Hz) compared to lines lists (1/7200~Hz) and is due to the needs of the \ac{SGWB} searches versus \ac{CW} searches.
Low amplitude, narrow lines will not contribute significant power in such coarse frequency bins.
Second, \ac{SGWB} searches are based on cross-correlation between spatially separated detectors, which suppresses uncorrelated instrumental noise.
Therefore, spectral artifacts that are not present in both detectors, or that do not exhibit correlated behavior, are further attenuated in the cross-correlation statistic.
In practice, this leads to a more compact set of excluded frequency bands compared to lines lists, while still effectively mitigating biases in the estimator for the \ac{SGWB} searches. 

Notch lists combine multiple sources of information from \ac{DetChar} studies. 
These include known instrumental features such as intentionally injected lines (e.g., calibration lines and simulated \ac{CW} signals), power mains harmonics, mechanical resonances of suspension and optical components, and electronic artifacts originating from cables and control systems. 
In addition, the notch lists also include empirically identified features, including prominent peaks and structures observed in \ac{CSD} between detectors, as well as statistically significant coherence outliers identified through monitoring tools such as StochMon (\sref{sec:Stochmon}) and coherence studies with auxiliary channels (\sref{sec:coherence} and \sref{sec:STAMP-PEM}).

Additionally, frequency bins that exhibit large temporal variance across \ac{PSD} estimates indicate non-stationary data or intermittent noise sources.
These frequency bins violate the stationary data assumption underlying cross-correlation analyses. 
Therefore, the identified frequency bins are cross-checked with the lines lists (\sref{sec:line-lists}) for instrumental origins and included in the notch list used by \ac{SGWB} search analyses.

\section{\label{sec:future}Discussion and conclusions}
We have described narrowband persistent spectral artifact monitoring and efforts to mitigate their impact on \ac{LIGO} $h(t)$ during the fourth \ac{LVK} observing run.
Several tools are used to monitor detector data, and the newly redesigned and reimplemented Fscan tool runs on daily, weekly, and monthly cadences.
Other tools to track specific artifacts rely on the data generated from Fscan during these production runs.
The complementary STAMP-PEM and StochMon tools are also used to monitor O4 data.
In order to maximize the benefit of limited resources during O4, we made use of a practical, physically motivated prioritization ordering of the artifacts that are most detrimental to narrowband persistent \ac{GW} searches.

Using these tools, we have discussed several case studies of narrowband persistent spectral artifacts, and, in some cases, how they were mitigated.
While progress was made during O4, sometimes a significant portion of the run was impacted in various ways by artifacts that were eventually mitigated.
We have attempted to simplify the sometimes bespoke investigation procedure for line and comb artifacts through the use of a common set of tools, thus improving efficiency.
It remains a challenge, however, to connect observed artifacts with noise sources in a timely fashion.
Contamination of data may continue for many months while the investigation and mitigation procedure unfolds.

We have discussed data quality products for narrowband persistent \ac{GW} searches: lines lists and notch lists.
These are crucial for searches to maximize their sensitivity to \acp{GW} and minimize costly follow-up of artifact-induced outliers.
Searches for persistent narrowband \ac{GW} signals are degraded even if a portion of the data is contaminated by lines in the frequency band of interest, so it is maximally beneficial to mitigate these problems early in data collection.
Once a line is vetted and included in data quality products such as lines lists or notch lists, the impacted frequency region is typically excluded from the whole analysis.
This means that some frequency bands of good data quality from later in the run may be vetoed, depending on the veto strategy of particular searches, degrading the search sensitivity (i.e., increasing the likelihood of false dismissal).

Since it is often unclear the cause of particular artifacts in $h(t)$ data, we advocate for thorough characterization of any new electronics in advance of installation and then comparing $h(t)$ spectral data before and after electronics are installed.
When it is not possible to do this on a device-by-device basis, for example when a completely new subsystem is installed, careful study of the noise is necessary.
Caution is advised when additional excitations are purposefully added to interferometer control loops (i.e., calibration lines, optical cavity alignment dither lines, etc.) because many more lines can be produced due to intermodulation problems.
On the whole, intermodulation problems affecting broad frequency bands have not been solved and need further investigation; perhaps utilizing bicoherence for additional insight~\cite{Hall_2024}.

Additional efforts are needed in order to automate more of the problem identification and investigation workflow, raise priority of line mitigation efforts, and speed the process of generating lines lists.
At present, these typically require bespoke investigation and vetting by scientists.
In a future era of continual data collection, we will be unable to spend time investigating historical data to identify artifacts.

Besides more automation efforts, interferometer scientists, engineers, and technicians must be vigilant to detrimental impacts of poorly characterized subsystems.
Electrical connections must be made with care in order to ensure that noise couplings to $h(t)$ are minimized.
We have found that maintaining only the required powered devices for low-noise, high-sensitivity operations be connected, and unplugging all other optional devices has proven to minimize the amount of narrowband persistent artifacts; \ac{LIGO} Livingston has taken this approach with success.

Noise subtraction methods of various forms~\citep{PhysRevD.110.122004,PhysRevD.101.042003,PhysRevD.105.102005,PhysRevResearch.2.033066,reissel2025coherencedeepcleanautonomousdenoising,zhou2026mitigationincoherentspectrallines} have been proposed as a possible solution for line artifacts, but further effort is needed to validate such mitigation efforts on a large scale and over long observing intervals.
Particularly loud lines, those which have few or no auxiliary channel witnesses, weak lines that require long measurement to accumulate enough signal above the background, lines created by non-linear coupling, or lines that experience discontinuous phase changes may prove especially challenging to mitigate such that a \ac{CW} signal could be reliably detected in the presence of such artifacts.

Candidate narrowband persistent \ac{GW} signals will require careful vetting in order to validate the credibility of such candidates.
Perhaps more challenging to validate are candidate signals that are persistent over a short epoch, so-called ``transient \ac{CW} signals''~\citep{RilesLivingReviewCW}.
As we have shown, characterizing non-astrophysical detector artifacts will play a key role in such validation efforts.
Building out a set of tools like those described here---Fscan, STAMP-PEM, and StochMon---is only one aspect of such efforts.
More effort is needed, however, in order to establish criteria and thresholds that may aid vetoing non-astrophysical candidates caused by noise artifacts or supporting the credibility of candidates and initiating more detailed investigations.
We anticipate further work in order to robustly determine these criteria.

\ack
The authors are grateful for the support of the \ac{LVK} collaboration Continuous Waves, Detector Characterization, and Stochastic working groups.
We thank the \ac{LVK} members that have participated in \ac{DetChar} and Stochastic group data quality shifts during O4.
Besides packages mentioned above, this work made use of the following software packages: \texttt{astropy}~\citep{astropy:2013,astropy:2018,astropy:2022,astropy_17756022}, \texttt{matplotlib}~\citep{Hunter:2007}, \texttt{numpy}~\citep{numpy}, and \texttt{python}~\citep{python}.
Software citation information aggregated using \texttt{\href{https://www.tomwagg.com/software-citation-station/}{The Software Citation Station}} \citep{software-citation-station-paper,software-citation-station-zenodo}.
This material is based upon work supported by the National Science Foundation’s LIGO Laboratory which is a major facility fully funded by the National Science Foundation.
The authors are grateful for computational resources provided by the LIGO Laboratory and supported by National Science Foundation Grants PHY-0757058 and PHY-0823459.
The authors also gratefully acknowledge research support in addition from
the Natural Sciences and Engineering Research Council of Canada (NSERC),
the Canadian Foundation for Innovation (CFI),
the Spanish Agencia Estatal de Investigaci\'on (AEI),
the Spanish Ministerio de Ciencia, Innovaci\'on y Universidades,
the Comunitat Auton\`oma de les Illes Balears through the Conselleria d'Educaci\'o i Universitats, Spain,
the Science and Technology Facilities Council (STFC) of the United Kingdom, 
the Max-Planck-Society (MPS), 
the State of Niedersachsen/Germany,
the Australian Research Council,
the French Centre National de la Recherche Scientifique (CNRS)
the Netherlands Organization for Scientific Research (NWO)
the Council of Scientific and Industrial Research of India, 
the Department of Science and Technology, India,
the Science \& Engineering Research Board (SERB), India,
the Ministry of Human Resource Development, India,
the CERCA Programme Generalitat de Catalunya, Spain,
the Polish National Agency for Academic Exchange,
the National Science Centre of Poland and the European Union - European Regional Development Fund;
the Foundation for Polish Science (FNP),
the Polish Ministry of Science and Higher Education,
the Swiss National Science Foundation (SNSF),
the European Commission,
the European Social Funds (ESF),
the European Regional Development Funds (ERDF),
the Royal Society, 
the Scottish Funding Council, 
the Scottish Universities Physics Alliance, 
the Belgian Fonds de la Recherche Scientifique (FRS-FNRS), 
Actions de Recherche Concert\'ees (ARC) and
Fonds Wetenschappelijk Onderzoek - Vlaanderen (FWO), Belgium,
the National Research Foundation of Korea,
the Brazilian Ministry of Science, Technology, and Innovations,
the International Center for Theoretical Physics South American Institute for Fundamental Research (ICTP-SAIFR), 
the National Natural Science Foundation of China (NSFC),
the National Science and Technology Council (NSTC), Taiwan,
the United States Department of Energy,
and
the Kavli Foundation.

\bibliography{references}

@article{Aasi2015,
doi = {10.1088/0264-9381/32/7/074001},
url = {https://doi.org/10.1088/0264-9381/32/7/074001},
year = {2015},
month = {mar},
publisher = {IOP Publishing},
volume = {32},
number = {7},
pages = {074001},
author = {Aasi, J and others},
title = {Advanced LIGO},
journal = {Classical and Quantum Gravity}
}

@article{Acernese_2015,
doi = {10.1088/0264-9381/32/2/024001},
url = {https://doi.org/10.1088/0264-9381/32/2/024001},
year = {2014},
month = {dec},
publisher = {IOP Publishing},
volume = {32},
number = {2},
pages = {024001},
author = {Acernese, F and others},
title = {Advanced Virgo: a second-generation interferometric gravitational wave detector},
journal = {Classical and Quantum Gravity}
}

@article{10.1093/ptep/ptaa125,
    author = {Akutsu, T and others},
    title = {Overview of KAGRA: Detector design and construction history},
    journal = {Progress of Theoretical and Experimental Physics},
    volume = {2021},
    number = {5},
    pages = {05A101},
    year = {2020},
    month = {08},
    issn = {2050-3911},
    doi = {10.1093/ptep/ptaa125},
    url = {https://doi.org/10.1093/ptep/ptaa125},
    eprint = {https://academic.oup.com/ptep/article-pdf/2021/5/05A101/37974994/ptaa125.pdf},
}

@article{Soni_2025,
doi = {10.1088/1361-6382/adc4b6},
url = {https://doi.org/10.1088/1361-6382/adc4b6},
year = {2025},
month = {apr},
publisher = {IOP Publishing},
volume = {42},
number = {8},
pages = {085016},
author = {Soni, S and others},
title = {LIGO Detector Characterization in the first half of the fourth Observing run},
journal = {Classical and Quantum Gravity}
}

@article{RilesLivingReviewCW,
	author = {Riles, Keith},
	date = {2023/04/17},
	id = {Riles2023},
	isbn = {1433-8351},
	journal = {Living Reviews in Relativity},
	number = {1},
	pages = {3},
	title = {Searches for continuous-wave gravitational radiation},
	volume = {26},
	year = {2023}
}

@article{PhysRevD.72.102004,
  title = {First all-sky upper limits from LIGO on the strength of periodic gravitational waves using the Hough transform},
  author = {Abbott, B. and others},
  collaboration = {LIGO Scientific Collaboration},
  journal = {Phys. Rev. D},
  volume = {72},
  issue = {10},
  pages = {102004},
  numpages = {22},
  year = {2005},
  month = {Nov},
  publisher = {American Physical Society},
  doi = {10.1103/PhysRevD.72.102004},
  url = {https://link.aps.org/doi/10.1103/PhysRevD.72.102004}
}

@article{PhysRevD.97.082002,
  title = {Identification and mitigation of narrow spectral artifacts that degrade searches for persistent gravitational waves in the first two observing runs of Advanced LIGO},
  author = {Covas, P. B. and others},
  collaboration = {LSC Instrument Authors},
  journal = {Phys. Rev. D},
  volume = {97},
  issue = {8},
  pages = {082002},
  numpages = {22},
  year = {2018},
  month = {Apr},
  publisher = {American Physical Society},
  doi = {10.1103/PhysRevD.97.082002},
  url = {https://link.aps.org/doi/10.1103/PhysRevD.97.082002}
}

@misc{goetz_2025_18776397,
  author       = {Goetz, Evan and
                  Neunzert, Ansel and
                  Suyamprakasam, Sudhagar},
  title        = {Fscan},
  month        = dec,
  year         = 2025,
  publisher    = {Zenodo},
  version      = {0.6.4},
  doi          = {10.5281/zenodo.18776397},
  url          = {https://doi.org/10.5281/zenodo.18776397},
}

@misc{htcondor_team_2024_14238998,
  author       = {HTCondor Team},
  title        = {HTCondor},
  month        = nov,
  year         = 2024,
  publisher    = {Zenodo},
  version      = {24.2.1},
  doi          = {10.5281/zenodo.14238998},
  url          = {https://doi.org/10.5281/zenodo.14238998},
}

@article{10.1063/1.4967303,
    author = {Karki, S. and others},
    title = {The Advanced LIGO photon calibrators},
    journal = {Review of Scientific Instruments},
    volume = {87},
    number = {11},
    pages = {114503},
    year = {2016},
    month = {11},
    issn = {0034-6748},
    doi = {10.1063/1.4967303},
    url = "{https://doi.org/10.1063/1.4967303}",
    eprint = "{https://pubs.aip.org/aip/rsi/article-pdf/doi/10.1063/1.4967303/15850013/114503_1_online.pdf}",
}

@misc{alog:66418,
       author = {Ansel Neunzert},
        title = "{Combs below 300 Hz in November / early December H1 data}",
 howpublished = "\url{https://alog.ligo-wa.caltech.edu/aLOG/index.php?callRep=66418}"
}

@misc{alog:66925,
       author = {Evan Goetz and others},
        title = "{Investigating H1 narrow spectral artifacts: Nov 2022 - mid-Jan 2023}",
 howpublished = "\url{https://alog.ligo-wa.caltech.edu/aLOG/index.php?callRep=66925}"
}

@misc{alog:69791,
       author = {Ansel Neunzert},
        title = "{1.6611 Hz comb disappeared between May 18 and May 19}",
 howpublished = "\url{https://alog.ligo-wa.caltech.edu/aLOG/index.php?callRep=69791}"
}

@misc{alog:71108,
       author = {Ansel Neunzert},
        title = "{1.6611 Hz comb re-appeared June 27}",
 howpublished = "\url{https://alog.ligo-wa.caltech.edu/aLOG/index.php?callRep=71108}"
}

@misc{alog:71533,
       author = {Victoriaa Xu},
        title = "{Collecting sqz / no-sqz times at hot, warm, cold, OM2 for quantum noise budgeting}",
 howpublished = "\url{https://alog.ligo-wa.caltech.edu/aLOG/index.php?callRep=71533}"
}

@misc{alog:71801,
       author = {Evan Goetz},
        title = "{OM2 heater suspected culprit for 1.6611 Hz comb in h(t)}",
 howpublished = "\url{https://alog.ligo-wa.caltech.edu/aLOG/index.php?callRep=71801}"
}

@misc{alog:72269,
       author = {Ansel Neunzert},
        title = "{1.6611 Hz comb gone after OM2 Beckhoff cable disconnected}",
 howpublished = "\url{https://alog.ligo-wa.caltech.edu/aLOG/index.php?callRep=72269}"
}

@misc{alog:68261,
       author = {Ansel Neunzert and Keit Pham},
        title = "{Persistent combs below 300 Hz in recent LHO data}",
 howpublished = "\url{https://alog.ligo-wa.caltech.edu/aLOG/index.php?callRep=68261}"
}

@misc{alog:74585,
       author = {Ansel Neunzert},
        title = "{Very weak 0.996785 Hz comb present for a few months}",
 howpublished = "\url{https://alog.ligo-wa.caltech.edu/aLOG/index.php?callRep=74585}"
}

@misc{alog:74617,
       author = {Ansel Neunzert and Camilla Compton},
        title = "{ITM HWS Camera Sync Frequencies Adjusted to 7Hz for DARM Comb Search}",
 howpublished = "\url{https://alog.ligo-wa.caltech.edu/aLOG/index.php?callRep=74617}"
}

@misc{T2100200,
       author = {Evan Goetz and others},
        title = "{O3 lines and combs in found in self-gated C01 data}",
 howpublished = "\url{https://dcc.ligo.org/LIGO-T2100200/public}"
}

@misc{alog:74750,
       author = {Camilla Compton},
        title = "{ITM HWS Camera's Powered off at 16:15UTC, search for cause of DARM Comb}",
 howpublished = "\url{https://alog.ligo-wa.caltech.edu/aLOG/index.php?callRep=74750}"
}

@misc{alog:74900,
       author = {Camilla Compton and Ansel Neunzert},
        title = "{7Hz comb gone in magnetometer channel after ITM HWS setting change}",
 howpublished = "\url{https://alog.ligo-wa.caltech.edu/aLOG/index.php?callRep=74900}"
}

@misc{alog:70219,
       author = {Ansel Neunzert},
        title = "{Combs contaminate many spectral bins in H1 weekly averaged data}",
 howpublished = "\url{https://alog.ligo-wa.caltech.edu/aLOG/index.php?callRep=70219}"
}

@misc{alog:70030,
       author = {Ansel Neunzert},
        title = "{near-9.5 Hz comb triplet spotted in L1 spectrum, similar to comb triplet at H1}",
 howpublished = "\url{https://alog.ligo-la.caltech.edu/aLOG/index.php?callRep=70030}"
}

@misc{alog:77990,
       author = {Alex Ritzie and Ansel Neunzert},
        title = "{9.5 Hz and 11 Hz combs seen in PSL-ISS\_PD(A,B)\_REL\_OUT\_DQ at LHO and LLO}",
 howpublished = "\url{https://alog.ligo-wa.caltech.edu/aLOG/index.php?callRep=77990}"
}

@misc{alog:78142,
       author = {Ansel Neunzert and Sheila Dwyer},
        title = "{9.5 Hz comb triplet likely started early Sept 2021 amid PSL upgrade work}",
 howpublished = "\url{https://alog.ligo-wa.caltech.edu/aLOG/index.php?callRep=78142}"
}

@misc{alog:79533,
       author = {Jason Oberling and Filiberto Clara},
        title = "{PSL 9.5Hz Comb Hunt (LHO WP 12031)}",
 howpublished = "\url{https://alog.ligo-wa.caltech.edu/aLOG/index.php?callRep=79533}"
}

@misc{alog:79593,
       author = {Ryan Short and Filiberto Clara},
        title = "{PSL 9.5Hz Comb Hunt (LHO WP 12031)}",
 howpublished = "\url{https://alog.ligo-wa.caltech.edu/aLOG/index.php?callRep=79593}"
}

@misc{alog:73376,
       author = {Josh Collins and Michael Laxen and Matthew Heinze},
        title = "{Transfer PSL Beckhoff to dedicated power supply (WP\#11816)}",
 howpublished = "\url{https://alog.ligo-la.caltech.edu/aLOG/index.php?callRep=73376}"
}

@misc{alog:71706,
       author = {Jeffrey Kissel and Thomas Shaffer and Louis Dartez},
        title = "{Turned ON CAL\_AWG\_LINES Extra Calibration Lines Again 2023-07-25 22:21:15 UTC}",
 howpublished = "\url{https://alog.ligo-wa.caltech.edu/aLOG/index.php?callRep=71706}"
}

@misc{alog:71964,
       author = {Ansel Neunzert},
        title = "{CAL\_AWG\_LINES extra calibration lines create many narrow artifacts below 100 Hz}",
 howpublished = "\url{https://alog.ligo-wa.caltech.edu/aLOG/index.php?callRep=71964}"
}

@misc{alog:79825,
       author = {Ansel Neunzert and others},
        title = "{Deeper study of a 2023 test shows calibration lines + violin modes = more line contamination than previously understood}",
 howpublished = "\url{https://alog.ligo-wa.caltech.edu/aLOG/index.php?callRep=79825}"
}

@misc{alog:82329,
       author = {Evan Goetz and Jeffrey Kissel and Louis Dartez},
        title = "{Evidence for in-band lines caused by aliasing of lines in high-sampled frequency DCPD ADC}",
 howpublished = "\url{https://alog.ligo-wa.caltech.edu/aLOG/index.php?callRep=82329}"
}

@article{ViolinMode,
  title = {GW150914: The Advanced LIGO Detectors in the Era of First Discoveries},
  author = {Abbott, B. P. and others},
  collaboration = {LIGO Scientific Collaboration and Virgo Collaboration},
  journal = {Phys. Rev. Lett.},
  volume = {116},
  issue = {13},
  pages = {131103},
  numpages = {12},
  year = {2016},
  month = {Mar},
  publisher = {American Physical Society},
  doi = {10.1103/PhysRevLett.116.131103},
  url = {https://link.aps.org/doi/10.1103/PhysRevLett.116.131103}
}

@misc{alog:72096,
    author = "Kissel, J.",
    title = "\textit{{Low-frequency Thermalization lines are OFF as of 2023-08-09 16:40 UTC
}}",
    howpublished = "\url{https://alog.ligo-wa.caltech.edu/aLOG/index.php?callRep=72096}"
}

@misc{alog:82375,
       author = {Jeffrey Kissel and Evan Goetz and Louis Dartez},
        title = "{Digital Anti-Aliasing Options for 524kHz OMC DCPD Path}",
 howpublished = "\url{https://alog.ligo-wa.caltech.edu/aLOG/index.php?callRep=82375}"
}

@misc{alog:82455,
       author = {Evan Goetz},
        title = "{Anti-aliasing filters doing a good job to remove high frequency artifacts}",
 howpublished = "\url{https://alog.ligo-wa.caltech.edu/aLOG/index.php?callRep=82455}"
}

@misc{alog:84136,
       author = {Kiet Pham and Sheila Dwyers and Ansel Neuzert},
        title = "{ Relationship between violin mode height and narrow spectral artifact contamination}",
 howpublished = "\url{https://alog.ligo-wa.caltech.edu/aLOG/index.php?callRep=84136}"
}

@misc{alog:82512,
       author = {Jeffrey Kissel},
        title = "{Confirmed with New Infrastructure: Last Week's Additional Digital AA Filtering Cleans Up Several Lines above 400 Hz}",
 howpublished = "\url{https://alog.ligo-wa.caltech.edu/aLOG/index.php?callRep=82512}"
}

@misc{alog:71501,
       author = {Taylor Starkman},
        title = "{Effect of violin mode amplitude on narrow lines near the violin modes}",
 howpublished = "\url{https://alog.ligo-wa.caltech.edu/aLOG/index.php?callRep=71501}"
}

@misc{alog:71800,
       author = {Taylor Starkman},
        title = "{ Asymmetry in narrow line contamination in bands around violin mode frequencies }",
 howpublished = "\url{https://alog.ligo-wa.caltech.edu/aLOG/index.php?callRep=71800}"
}

@misc{alog:82320,
       author = {Ansel Neunzert},
        title = "{Relationship between violin mode height and narrow spectral artifact contamination, revisited}",
 howpublished = "\url{https://alog.ligo-wa.caltech.edu/aLOG/index.php?callRep=82320}"
}

@misc{alog:87923,
       author = {Alicia Calafat and Joan-Rene Merou and Sheila Dwyer and Robert Schofield and Jenne Driggers},
        title = "{Intermodulation lines and $\beta$ estimation in H1:OMC-DCPD\_SUM\_OUT\_DQ}",
 howpublished = "\url{https://alog.ligo-wa.caltech.edu/aLOG/index.php?callRep=87923}"
}

@misc{alog:87414,
       author = {Joan-Rene Merou and others},
        title = "{Witness channels of the set of near-30Hz and near-100Hz combs at LHO}",
 howpublished = "\url{https://alog.ligo-wa.caltech.edu/aLOG/index.php?callRep=87414}"
}

@misc{alog:87889,
       author = {Joan-Rene Merou and Alicia Calafat and Sheila Dwyer and Robert Schofield},
        title = "{Correlation between ITMX and ITMY ESD bias settings and near-30 Hz near-100 Hz line amplitudes}",
 howpublished = "\url{https://alog.ligo-wa.caltech.edu/aLOG/index.php?callRep=87889}"
}

@misc{alog:88089,
       author = {Joan-Rene Merou and Alicia Calafat and Sheila Dwyer and Jenne Driggers},
        title = "{With voltage set to 0, grounding of the ITMX ESD driver}",
 howpublished = "\url{https://alog.ligo-wa.caltech.edu/aLOG/index.php?callRep=88089}"
}

@misc{alog:88595,
       author = {Joan-Rene Merou and Alicia Calafat and Sheila Dwyer and Robert Schofield and Anamaria Effler},
        title = "{Evidence linking PSL video cameras to the near-30 Hz (29.97 Hz) comb observed in DARM}",
 howpublished = "\url{https://alog.ligo-wa.caltech.edu/aLOG/index.php?callRep=88595}"
}

@techreport{axis215ptz_datasheet,
  author       = {{Axis Communications}},
  title        = {{AXIS 215 PTZ Network Camera Datasheet}},
  institution  = {Axis Communications},
  year         = {2009},
  number       = {34462},
  url          = {https://www.axis.com/files/datasheet/ds_215ptz_34462_en_0902_lo.pdf},
  note         = {Accessed: 2026-04-15}
}

@misc{NTCSstandard,
  author = {{National Television System Committee}},
  title  = {NTSC Television Standard},
  year   = {1953},
  note   = {Color NTSC operates at approximately 29.97 frames per second due to color subcarrier constraints},
  url    = {https://www.itu.int/rec/R-REC-BT.470/en}
}

@misc{alog:66036,
       author = {Joseph Bayley and others},
        title = "{Odd harmonics of an 11.904 Hz comb present in both L1 and H1}",
 howpublished = "\url{https://alog.ligo-la.caltech.edu/aLOG/index.php?callRep=66036}"
}

@misc{alog:72866,
       author = {Ansel Neunzert},
        title = "{History of the 11.9 Hz comb in DARM}",
 howpublished = "\url{https://alog.ligo-la.caltech.edu/aLOG/index.php?callRep=72866}"
}

@misc{alog:66385,
       author = {Joseph Betzieser and Michae Laxen and Anamaria Effler},
        title = "{EX Beckhoff ON/OFF test for locating spurious 11.9 Hz comb}",
 howpublished = "\url{https://alog.ligo-la.caltech.edu/aLOG/index.php?callRep=66385}"
}

@misc{alog:67581,
       author = {Anamaria Effler and Tabata Ferreira},
        title = "{Disappearance of the 35.7Hz line on the DARM channel}",
 howpublished = "\url{https://alog.ligo-la.caltech.edu/aLOG/index.php?callRep=67581}"
}

@article{PhysRevD.108.022003,
  title = {Timing system of LIGO discoveries},
  author = {Sullivan, Andrew G. and others},
  journal = {Phys. Rev. D},
  volume = {108},
  issue = {2},
  pages = {022003},
  numpages = {13},
  year = {2023},
  month = {Jul},
  publisher = {American Physical Society},
  doi = {10.1103/PhysRevD.108.022003},
  url = {https://link.aps.org/doi/10.1103/PhysRevD.108.022003}
}

@misc{alog:64525,
       author = {Shivaraj Kandhasamy and others},
        title = "{DuoTone line frequencies in DARM spectra}",
 howpublished = "\url{https://alog.ligo-la.caltech.edu/aLOG/index.php?callRep=64525}"
}

@misc{alog:77696,
       author = {Joseph Betzwieser},
        title = "{DuoTone signal seen in h(t)}",
 howpublished = "\url{https://alog.ligo-wa.caltech.edu/aLOG/index.php?callRep=77696}"
}

@misc{alog:82647,
       author = {David Barker and others},
        title = "{CDS Maintenance Summary: Tuesday 4th February 2025}",
 howpublished = "\url{https://alog.ligo-wa.caltech.edu/aLOG/index.php?callRep=82647}"
}

@misc{alog:72103,
       author = {Caleb Wheeler and others},
        title = "{WP \#12045 - DuoTone upgrade}",
 howpublished = "\url{https://alog.ligo-la.caltech.edu/aLOG/index.php?callRep=75103}"
}

@phdthesis{Meyers:2018nyo,
    author = "Meyers, Patrick Michael",
    title = "{Cross-correlation searches for persistent gravitational waves with Advanced {LIGO} and noise studies for current and future ground-based gravitational-wave detectors}",
    school = "Minnesota U.",
    year = "2018"
}

@techreport{Stochmon,
      author         = "Hernandez, G and Thrane. E",
      title          = "\textit{{Stochmon: A LIGO Data Analysis Tool}}",
      number         = "T1400205",
      year           = "2023",
      institution    = {LSC},
      url            = {https://dcc.ligo.org/LIGO-T1400205/public}
}

@article{Hall_2024,
doi = {10.1088/1361-6382/ad7cb7},
url = {https://doi.org/10.1088/1361-6382/ad7cb7},
year = {2024},
month = {nov},
publisher = {IOP Publishing},
volume = {41},
number = {24},
pages = {245016},
author = {Hall, Bernard and Suyamprakasam, Sudhagar and Mazumder, Nairwita and More, Anupreeta and Bose, Sukanta},
title = {Identifying noise transients in gravitational-wave data arising from nonlinear couplings},
journal = {Classical and Quantum Gravity}
}

@article{Coughlin_2010,
doi = {10.1088/1742-6596/243/1/012010},
url = {https://doi.org/10.1088/1742-6596/243/1/012010},
year = {2010},
month = {nov},
publisher = {},
volume = {243},
number = {1},
pages = {012010},
author = {Michael Coughlin and (for the Ligo Scientific Collaboration and the Virgo Collaboration)},
title = {Noise Line Identification in LIGO S6 and Virgo VSR2},
journal = {Journal of Physics: Conference Series}
}

@misc{T2400204,
       author = {Evan Goetz and others},
        title = "{O4a lines and combs in found in self-gated C00 cleaned data}",
 howpublished = "\url{https://dcc.ligo.org/LIGO-T2400204/public}"
}

@misc{enwiki:1351281889,
    author = "{Wikipedia contributors}",
    title = "Z-test --- {Wikipedia}{,} The Free Encyclopedia",
    year = "2026",
    howpublished = "\url{https://en.wikipedia.org/w/index.php?title=Z-test&oldid=1351281889}",
    note = "[Online; accessed 1-May-2026]"
  }

@misc{astropy_17756022,
  author       = {Astropy Collaboration},
  title        = {Astropy},
  month        = nov,
  year         = 2025,
  publisher    = {Zenodo},
  version      = {v7.2.0},
  doi          = {10.5281/zenodo.17756022},
  howpublished          = "\url{https://doi.org/10.5281/zenodo.17756022}",
}

@article{astropy:2013,
  adsurl        = {http://adsabs.harvard.edu/abs/2013A\%26A...558A..33A},
  archiveprefix = {arXiv},
  author        = {{Astropy Collaboration} and others},
  doi           = {10.1051/0004-6361/201322068},
  eid           = {A33},
  eprint        = {1307.6212},
  journal       = {Astronomy and Astrophysics},
  month         = oct,
  pages         = {A33},
  primaryclass  = {astro-ph.IM},
  title         = {{Astropy: A community Python package for astronomy}},
  volume        = 558,
  year          = 2013
}

@article{astropy:2018,
  author        = {{Astropy Collaboration} and {Astropy Contributors}},
  title         = "{The Astropy Project: Building an Open-science Project and Status of the v2.0 Core Package}",
  journal       = {The Astronomical Journal},
  year          = 2018,
  month         = sep,
  volume        = {156},
  number        = {3},
  eid           = {123},
  pages         = {123},
  doi           = {10.3847/1538-3881/aabc4f},
  archiveprefix = {arXiv},
  eprint        = {1801.02634},
  primaryclass  = {astro-ph.IM},
  adsurl        = {https://ui.adsabs.harvard.edu/abs/2018AJ....156..123A}
}

@article{astropy:2022,
  author        = {{Astropy Collaboration} and {Astropy Project Contributors}},
  title         = "{The Astropy Project: Sustaining and Growing a Community-oriented Open-source Project and the Latest Major Release (v5.0) of the Core Package}",
  journal       = {The Astrophysical Journal},
  year          = 2022,
  month         = aug,
  volume        = {935},
  number        = {2},
  eid           = {167},
  pages         = {167},
  doi           = {10.3847/1538-4357/ac7c74},
  archiveprefix = {arXiv},
  eprint        = {2206.14220},
  primaryclass  = {astro-ph.IM},
  adsurl        = {https://ui.adsabs.harvard.edu/abs/2022ApJ...935..167A}
}

@article{Hunter:2007,
  author        = {Hunter, J. D.},
  title         = {Matplotlib: A 2D graphics environment},
  journal       = {Computing in Science \& Engineering},
  volume        = {9},
  number        = {3},
  pages         = {90--95},
  publisher     = {IEEE COMPUTER SOC},
  doi           = {10.1109/MCSE.2007.55},
  year          = 2007
}

@article{numpy,
  title         = {Array programming with {NumPy}},
  author        = {Charles R. Harris and others},
  year          = {2020},
  month         = sep,
  journal       = {Nature},
  volume        = {585},
  number        = {7825},
  pages         = {357--362},
  doi           = {10.1038/s41586-020-2649-2},
  publisher     = {Springer Science and Business Media {LLC}},
  howpublished           = "\url{https://doi.org/10.1038/s41586-020-2649-2}"
}

@book{python,
  author        = {Van Rossum, Guido and Drake, Fred L.},
  title         = {Python 3 Reference Manual},
  year          = {2009},
  isbn          = {1441412697},
  publisher     = {CreateSpace},
  address       = {Scotts Valley, CA}
}

@misc{scipy_18736568,
  author       = {Ralf Gommers and
                  others},
  title        = {scipy/scipy: SciPy 1.17.1},
  month        = feb,
  year         = 2026,
  publisher    = {Zenodo},
  version      = {v1.17.1},
  doi          = {10.5281/zenodo.18736568},
  howpublished          = "\url{https://doi.org/10.5281/zenodo.18736568}",
  swhid        = {swh:1:dir:934360229a7c597c39d811831bb6c020ef7f8151
                   ;origin=https://doi.org/10.5281/zenodo.595738;visi
                   t=swh:1:snp:2125cfcb4a989632c0e87df14bff9dc83bc816
                   e7;anchor=swh:1:rel:66e87815fdd0136976a21a8a0284e5
                   5c4e5b156e;path=scipy-scipy-c59ed93
                  },
}

@article{2020SciPy-NMeth,
  author        = {Virtanen, Pauli and others and {SciPy 1.0 Contributors}},
  title         = {{{SciPy} 1.0: Fundamental Algorithms for Scientific Computing in Python}},
  journal       = {Nature Methods},
  year          = {2020},
  volume        = {17},
  pages         = {261--272},
  adsurl        = {https://rdcu.be/b08Wh},
  doi           = {10.1038/s41592-019-0686-2}
}

@misc{Bokeh,
  title         = {Bokeh: Python library for interactive visualization},
  author        = {{Bokeh Development Team}},
  year          = {2018},
  howpublished           = "\url{https://bokeh.pydata.org/en/latest/}"
}

@misc{lalsuite,
  author        = "{LIGO Scientific Collaboration} and {Virgo Collaboration} and {KAGRA Collaboration}",
  title         = "{LVK} {A}lgorithm {L}ibrary - {LALS}uite",
  howpublished  = "Free software (GPL)",
  doi           = "10.7935/GT1W-FZ16",
  year          = "2018"
}

@article{software-citation-station-paper,
  author        = {{Wagg}, Tom and {Broekgaarden}, Floor S.},
  title         = "{Streamlining and standardizing software citations with The Software Citation Station}",
  journal       = {arXiv e-prints},
  year          = 2024,
  month         = jun,
  eid           = {arXiv:2406.04405},
  pages         = {arXiv:2406.04405},
  archiveprefix = {arXiv},
  eprint        = {2406.04405},
  primaryclass  = {astro-ph.IM},
  adsurl        = {https://ui.adsabs.harvard.edu/abs/2024arXiv240604405W}
}

@misc{software-citation-station-zenodo,
  author       = {Tom Wagg and
                  Floor Broekgaarden and
                  Phil Van-Lane and
                  Kai Wu and
                  Kayhan Gültekin},
  title        = {TomWagg/software-citation-station: v1.4},
  month        = nov,
  year         = 2025,
  publisher    = {Zenodo},
  version      = {v1.4},
  doi          = {10.5281/zenodo.17654855},
  howpublished          = "\url{https://doi.org/10.5281/zenodo.17654855}",
  swhid        = {swh:1:dir:9a009430037c791424a572f542e9a5d5c1fb44ff
                   ;origin=https://doi.org/10.5281/zenodo.13225526;vi
                   sit=swh:1:snp:ef11f058d718d691f0661c9445c2251328cb
                   ac95;anchor=swh:1:rel:84cde4e532032537b7f014c4467c
                   102430a53fa2;path=TomWagg-software-citation-
                   station-61a588a
                  },
}

@misc{o4a_allsky_cw,
      title={All-sky Searches for Continuous Gravitational Waves from Isolated Neutron Stars in the Data from the First Part of the Fourth LIGO-Virgo-KAGRA Observing Run}, 
      author={The LIGO Scientific Collaboration and the Virgo Collaboration and the KAGRA Collaboration and others},
      year={2026},
      eprint={2603.14168},
      archivePrefix={arXiv},
      primaryClass={gr-qc},
      url={https://arxiv.org/abs/2603.14168}, 
}

@misc{T2400003,
       author = {Derek Davis and Ansel Neunzert and others},
        title = "{Self-gating of O4a h(t) for use in continuous-wave searches}",
 howpublished = "\url{https://dcc.ligo.org/LIGO-T2400003/public}"
}

@article{PhysRevD.110.122004,
  title = {Adaptive cancellation of mains power interference in continuous gravitational wave searches with a hidden Markov model},
  author = {Kimpson, T. and others},
  journal = {Phys. Rev. D},
  volume = {110},
  issue = {12},
  pages = {122004},
  numpages = {16},
  year = {2024},
  month = {Dec},
  publisher = {American Physical Society},
  doi = {10.1103/PhysRevD.110.122004},
  url = {https://link.aps.org/doi/10.1103/PhysRevD.110.122004}
}

@article{PhysRevD.101.042003,
  title = {Machine-learning nonstationary noise out of gravitational-wave detectors},
  author = {Vajente, G. and Huang, Y. and Isi, M. and Driggers, J. C. and Kissel, J. S. and Szczepa\ifmmode \acute{n}\else \'{n}\fi{}czyk, M. J. and Vitale, S.},
  journal = {Phys. Rev. D},
  volume = {101},
  issue = {4},
  pages = {042003},
  numpages = {12},
  year = {2020},
  month = {Feb},
  publisher = {American Physical Society},
  doi = {10.1103/PhysRevD.101.042003},
  url = {https://link.aps.org/doi/10.1103/PhysRevD.101.042003}
}

@article{PhysRevD.105.102005,
  title = {Data mining and machine learning improve gravitational-wave detector sensitivity},
  author = {Vajente, Gabriele},
  journal = {Phys. Rev. D},
  volume = {105},
  issue = {10},
  pages = {102005},
  numpages = {9},
  year = {2022},
  month = {May},
  publisher = {American Physical Society},
  doi = {10.1103/PhysRevD.105.102005},
  url = {https://link.aps.org/doi/10.1103/PhysRevD.105.102005}
}

@article{PhysRevResearch.2.033066,
  title = {Noise reduction in gravitational-wave data via deep learning},
  author = {Ormiston, Rich and Nguyen, Tri and Coughlin, Michael and Adhikari, Rana X. and Katsavounidis, Erik},
  journal = {Phys. Rev. Res.},
  volume = {2},
  issue = {3},
  pages = {033066},
  numpages = {11},
  year = {2020},
  month = {Jul},
  publisher = {American Physical Society},
  doi = {10.1103/PhysRevResearch.2.033066},
  url = {https://link.aps.org/doi/10.1103/PhysRevResearch.2.033066}
}

@misc{reissel2025coherencedeepcleanautonomousdenoising,
      title={Coherence DeepClean: Toward autonomous denoising of gravitational-wave detector data}, 
      author={Christina Reissel and Siddharth Soni and Muhammed Saleem and Michael Coughlin and Philip Harris and Erik Katsavounidis},
      year={2025},
      eprint={2501.04883},
      archivePrefix={arXiv},
      primaryClass={gr-qc},
      url={https://arxiv.org/abs/2501.04883}, 
}

@misc{zhou2026mitigationincoherentspectrallines,
      title={Mitigation of Incoherent Spectral Lines via Adaptive Coherence Analysis for Continuous Gravitational-Wave Searches}, 
      author={Ye Zhou and Karl Wette},
      year={2026},
      eprint={2604.01919},
      archivePrefix={arXiv},
      primaryClass={gr-qc},
      url={https://arxiv.org/abs/2604.01919}, 
}

\end{document}